\documentclass[twocolumn]{aastex631}
\usepackage{lineno}

\newcommand{\bs}{$\langle B \rangle$}

\accepted{for publication in ApJ, January 01, 2025}

\begin{document}
 

\title{Multi-Observatory Study of Young Stellar Energetic Flares (MORYSEF): No Evidence For Abnormally Strong Stellar Magnetic Fields After Powerful X-ray Flares}

\correspondingauthor{Konstantin Getman}
\email{kug1@psu.edu}

\author[0000-0002-6137-8280]{Konstantin V. Getman}
\affiliation{Department of Astronomy \& Astrophysics \\
Pennsylvania State University \\ 
525 Davey Laboratory \\
University Park, PA 16802, USA}

\author[0000-0003-3061-4591]{Oleg Kochukhov}
\affiliation{Department of Physics and Astronomy, Uppsala University, Box 516, 75120 Uppsala, Sweden}

\author[0000-0001-8720-5612]{Joe P. Ninan}
\affiliation{Department of Astronomy and Astrophysics, Tata Institute of Fundamental Research, Homi Bhabha Road, Colaba, Mumbai 400005, India}

\author[0000-0002-5077-6734]{Eric D. Feigelson}
\affiliation{Department of Astronomy \& Astrophysics \\
Pennsylvania State University \\ 
525 Davey Laboratory \\
University Park, PA 16802, USA}

\author[0000-0003-4452-0588]{Vladimir S. Airapetian}
\affiliation{American University, 4400 Massachusetts Avenue NW, Washington, DC 20016, USA USA}
\affiliation{NASA/GSFC/SEEC, Greenbelt, MD 20771, USA}

\author[0000-0002-1566-389X]{Abygail R. Waggoner}
\affiliation{University of Virginia, Charlottesville, VA 22904, USA}
\affiliation{Department of Astronomy, University of Wisconsin-Madison, 475 N Charter St, Madison, WI 53706}

\author[0000-0003-2076-8001]{L. Ilsedore Cleeves}
\affiliation{University of Virginia, Charlottesville, VA 22904, USA}

\author[0000-0001-8694-4966]{Jan Forbrich}
\affiliation{Centre for Astrophysics Research, University of Hertfordshire, College Lane, Hatfield, AL10 9AB, UK}

\author[0000-0001-6010-6200]{Sergio A. Dzib}
\affiliation{Max-Planck-Institut fur Radioastronomie (MPIfR), Auf dem Hugel 69, 53121 Bonn, Germany}

\author[0000-0003-1413-1776]{Charles J. Law}
\altaffiliation{NASA Hubble Fellowship Program Sagan Fellow}
\affiliation{University of Virginia, Charlottesville, VA 22904, USA}

\author[0000-0003-1817-6576]{Christian Rab}
\affiliation{University Observatory, Faculty of Physics, Ludwig-Maximilians-Universitat Munchen, Scheinerstr. 1, D-81679 Munich, Germany}
\affiliation{Max-Planck-Institut für extraterrestrische Physik, Giessenbachstrasse 1, D-85748 Garching, Germany}

\author[0000-0001-9626-0613]{Daniel M. Krolikowski}
\affiliation{Steward Observatory, The University of Arizona, 933 N. Cherry Ave, Tucson, AZ 85721, USA}

\begin{abstract}
We explore the empirical power-law relationship between X-ray luminosity ($L_X$) and total surface magnetic flux ($\Phi$), established across solar magnetic elements, time- and disk-averaged emission from the Sun, older active stars, and pre-main-sequence (PMS) stars. Previous models of large PMS X-ray flares, lacking direct magnetic field measurements, showed discrepancies from this baseline law, which MHD simulations attribute to unusually strong magnetic fields during flares. To test this, we used nearly simultaneous Chandra X-ray and HET-HPF near-infrared observations of four young Orion stars, measuring surface magnetic fields during or just after powerful PMS X-ray flares. We also modeled these PMS X-ray flares, incorporating their measured magnetic field strengths. Our findings reveal magnetic field strengths at the stellar surface typical of non-flaring PMS stars, ruling out the need for abnormally strong fields during flares. Both PMS and solar flares deviate from the $L_X - \Phi$ law, with PMS flares exhibiting a more pronounced deviation, primarily due to their much larger active regions on the surface and larger flaring loop volumes above the surface compared to their solar counterparts. These deviations likely stem from the fact that powerful flares are driven by magnetic reconnection, while baseline X-ray emission may involve less efficient mechanisms like Alfv\'{e}n wave heating. Our results also indicate a preference for dipolar magnetic loops in PMS flares, consistent with Zeeman-Doppler imaging of fully convective stars. This requirement for giant dipolar loops aligns with MHD predictions of strong dipoles supported by polar magnetic surface active regions in fast-rotating, fully convective stars.
\end{abstract}
 

\section{Introduction: Hypothesis of Exceptionally Strong Magnetic Fields Linked to Large PMS Flares} \label{sec:intro}

Stars exhibit their highest levels of X-ray activity during the fully convective pre-main sequence (PMS) phase \citep{Preibisch05}. PMS X-ray flares can be extraordinary: they are 3 to 6 orders of magnitude more powerful than the solar Carrington event \citep{Hudson2021}, occur 6 orders of magnitude more frequently, last 2 orders of magnitude longer, and span 2 to 3 orders of magnitude larger X-ray coronal flaring structures than those on the contemporary Sun \citep{Wolk05, Favata2005, Colombo07, Getman2021, Getman2021b}.

\begin{figure}[h]
    \centering
     \includegraphics[width=0.48\textwidth]{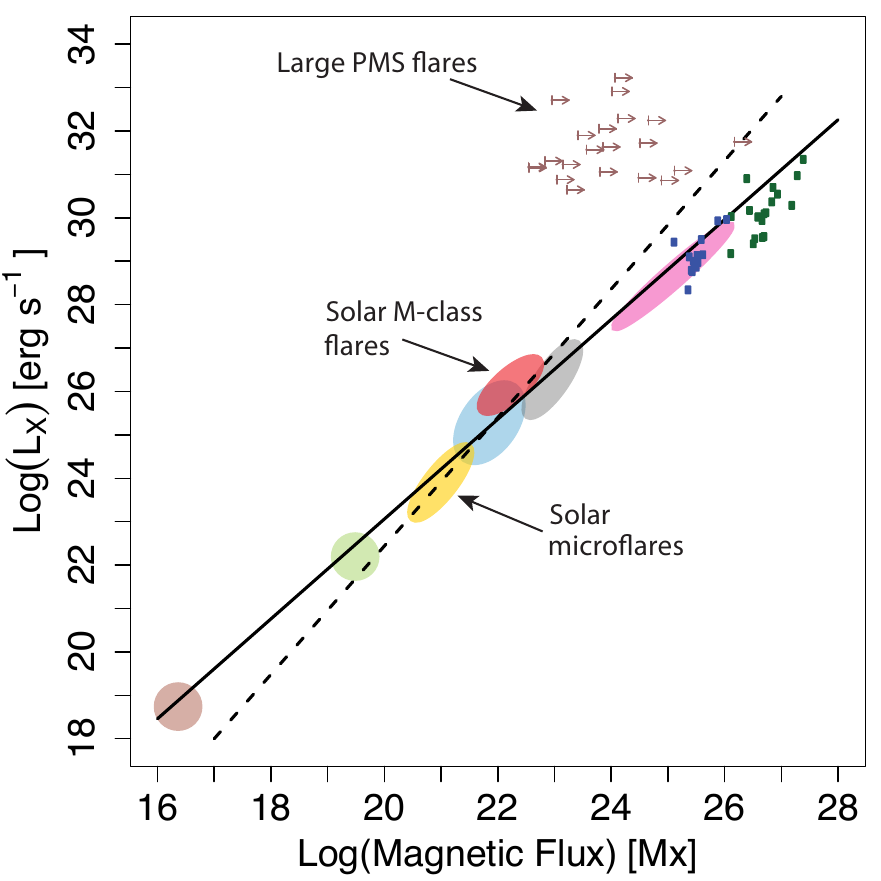}
    \caption{X-ray luminosity as a function of magnetic flux. Lower limits on the magnetic flux for PMS super-flares and mega-flares from \citet{Getman2021b} are indicated by the brown arrows.  The solid and dashed black lines are the relations $L_X \propto \Phi^{m}$ with $m=1.15$ \citep{Pevtsov2003} and $m=1.48$ \citep{Kirichenko2017}, respectively. These lines fit various magnetic elements emitting X-rays.  From \citet{Pevtsov2003}, these include the quiet Sun (brown), solar X-ray bright points (light green),  solar active regions (cyan), solar disk averages (grey), and old G, K, M dwarfs (magenta). Solar microflares (golden oval) are from \citet{Kirichenko2017} and solar M-class flares (red oval) are from \citet{Su2007}. Individual solar-type dwarfs (blue points) and Orion/Taurus PMS stars (dark green) with direct measurements of surface magnetic fields are from \citet{Kochukhov2020} and \citet{Sokal2020}, respectively.} \label{fig:pevtsov_plot}
\end{figure}

Both PMS and solar flares are driven by coronal magnetic reconnection processes, and substantial empirical evidence suggests that protoplanetary disks surrounding many young stars do not play a significant role in producing PMS X-ray super-flares ($10^{34} < E_X < 10^{36}$~erg) or mega-flares ($E_X > 10^{36}$~erg) \citep{Getman08b,Getman2021b,Getman2024}. While still undergoing gravitational contraction, solar-mass PMS stars have volumes 10 to 30 times larger than that of the contemporary Sun \citep{Bressan12,Chen14}. The convection-driven dynamo in PMS stars is significantly more efficient because it operates throughout these much larger volumes, unlike the Sun's dynamo, which is confined to the tachocline and convection zone \citep{Browning2008, Christensen2009, Kapyla2023}. The surface magnetic fluxes generated by the convection-driven dynamo in PMS stars, likely independent of stellar rotation and constrained only by the vast kinetic energy available in stellar convective flows \citep{Reiners2022}, are significantly stronger than those of the contemporary Sun. This results in much larger active regions and far more powerful X-ray flares.

\begin{figure*}
\centering
\includegraphics[width=0.95\textwidth]{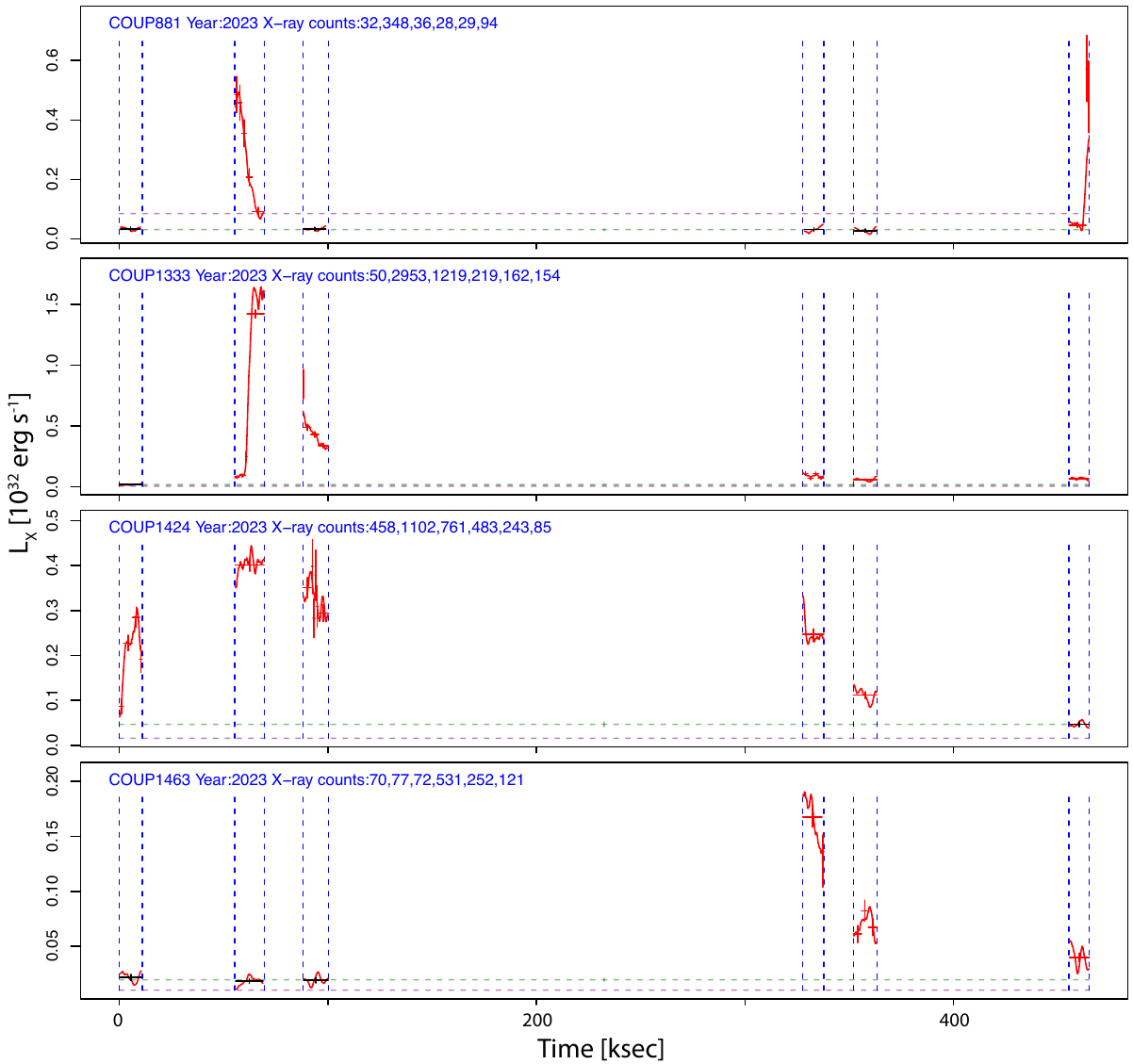}
\caption{{\it Chandra} X-ray light curves of the four HET-HPF stars are shown in units of intrinsic stellar X-ray luminosity. These lightcurves were derived in \citet{Getman2024}. The zero time point on the graph corresponds to an MJD value of 60295.20 days. Blue dashed lines indicate the start and stop times of each of the six {\it Chandra} observations. Kernel density estimation (KDE) curves, shown in red, are included solely as visual aids to suggest potential trends. Red and black Bayesian Block segments with error bars represent flare events and characteristic (i.e., baseline or quiescent) X-ray emission levels, respectively. The characteristic emission levels are also indicated by dashed lines: magenta for earlier epoch-2003 {\it Chandra} observations (as described in \citet{Getman2024}) and green for the 2023 epoch. Standard 1-$\sigma$ confidence intervals (marked by error bars) for both Bayesian Blocks and characteristic levels are calculated using the \citet{Gehrels86} statistic. The figure legends display the observation epoch and the total number of X-ray photons per observation.} \label{fig:chandra_lc}
\end{figure*}

Figure~\ref{fig:pevtsov_plot} demonstrates a universal power-law relationship between X-ray luminosity $L_X$ and total (unsigned) surface magnetic flux $\Phi = B \times A$, where $A$ is the surface area through which the magnetic field of strength $B$ passes. This empirical $L_X - \Phi$ relationship applies across solar magnetic elements, the Sun, older active stars, and PMS stars \citep{Pevtsov2003,Kirichenko2017} and suggests that the propagation and dissipation of magnetic fluxes follow a universal pattern, regardless of the nature of the underlying dynamos. However, notably, it also shows that the time-averaged X-ray luminosities and surface magnetic fluxes of PMS stars (dark green points) are 4-5 orders of magnitude higher than those of the contemporary Sun (grey oval), underscoring the aforementioned greater efficiency of PMS convection-driven dynamos.

A notable discrepancy is observed with the most powerful PMS super- and mega-flares, which have X-ray energies exceeding $E_X > 10^{35}$ erg (indicated by brown arrows in Figure~\ref{fig:pevtsov_plot}). The occurrence rate of such high-energy flares ($E_X > 10^{35}$ erg) in solar-mass PMS stars is approximately 20 per star per year \citep{Getman2021}. For these events, X-ray flare modeling --- assuming magnetic equipartition in the flaring coronal loops \citep{Getman2021b}, {\it and not involving direct magnetic field measurements} --- predicts that the observed $L_X$ values are $10^3-10^4$ times greater than those forecasted by scaling relations. 

The slope $m$ represents the efficiency of converting surface magnetic flux into X-ray luminosity. Slope $m > 1$ implies enhanced magnetic heating efficiency. Recent 3D-MHD simulations suggest a steeper slope, $m = 3.4$, in the $L_X \propto \Phi^m$ relationship for high-energy flares. However, this would necessitate magnetic field strengths in active regions reaching $B_{\text{spot}} \simeq (10-20)$ kG \citep{Zhuleku2021}. Consequently, this suggests the hypothesis that the surface-averaged magnetic field strength $\langle B \rangle = B_{\text{spot}} \times f$ --- where $f$ is the surface filling factor of magnetic spots, typically $f > (0.7-0.9)$ in fully convective PMS stars --- could be significantly higher during super- and mega-flares than during the more common, non-super-flaring PMS epochs, when $\langle B \rangle \sim (1-4)$ kG \citep{Sokal2020}.

The main goal of the current study is to test the above hypothesis by directly measuring the average surface magnetic fields of a few PMS stars immediately after these stars underwent super- or mega-flaring X-ray events. 

In December 2023, a related observational campaign was carried out as part of the extensive Multi-Observatory Research of Young Stellar Energetic Flares (MORYSEF) project, which seeks to investigate multiple aspects of PMS X-ray emission \citep{Getman2024}. The campaign utilized the {\it Chandra} X-ray Observatory, Atacama Large Millimeter/submillimeter Array (ALMA), Very Long Baseline Array (VLBA), and the Habitable-zone Planet Finder (HPF) near-infrared instrument on the Hobby-Eberly Telescope (HET). These facilities were employed to achieve the broad goals of studying particle ejections, disk ionization, and surface magnetic fields following powerful X-ray flares.

Specifically, for the purposes of this paper, the {\it Chandra} X-ray Observatory carried out six relatively short observations of the Orion Nebula star-forming region, which hosts several thousand fully convective PMS stars. {\it Chandra} detected dozens of super- and mega-flaring stars. Follow-up observations of four of these stars (COUP~881, COUP~1333, COUP~1424, COUP~1463) were conducted using HET-HPF \citep{Mahadevan2012,Mahadevan2014}. Surface magnetic field strengths for these stars were estimated through Zeeman broadening (ZB) measurements of  magnetically-sensitive NIR spectral lines \citep{Kochukhov2021}.

Our analysis shows that the average surface magnetic field strengths for the four investigated PMS stars in the Orion Nebula region, measured during or immediately after their powerful, long-duration X-ray flares, remain consistent with values typical of non-flaring PMS stars. These direct magnetic field measurements allow us to constrain the positions of large PMS X-ray flares on the $L_X - \Phi$ diagram and to investigate the origins of this phenomenon.

The paper is organized as follows. Section \ref{sec:observations} details the {\it Chandra} and HET-HPF observations, data reduction, and the sample of flaring PMS stars analyzed in this study. Section \ref{sec:ZB_mesurements} presents the magnetic field measurements for our four flaring PMS stars. Section \ref{sec:results} focuses on various scientific analyses, including a comparison with the magnetic fields of non-flaring stars, calculations of flaring loop cross-sections and free-to-potential magnetic energy ratios, as well as an examination of the inferred positions of powerful PMS and solar flares on the $L_X-\Phi$ plane. The underlying physical mechanisms responsible for the deviation of large X-ray flares from the baseline $L_X-\Phi$ relationship, along with the magnetic loop configurations associated with large PMS flares, are further discussed in Section~\ref{sec:discussion}. Section \ref{sec:conclusions} summarizes our findings.

\section{X-ray and NIR observations and data extraction} \label{sec:observations}
\subsection{{\it Chandra} observations and properties of four flaring PMS stars} \label{sec:chandra_observations}

The six {\it Chandra}-ACIS-I \citep{Weisskopf2002,Garmire2003} observations from December 2023, totaling an exposure of 67~ksec, along with the X-ray data reduction procedures and the analyses used to identify large X-ray flares and measure their energetics, are thoroughly detailed in \citet{Getman2024}. Each observation, lasting approximately $10–14$~ksec, was conducted between December 17 and December 22, 2023, with gaps between observations ranging from 10 to 220~ksec. 

In addition, for scientific purposes related to the multi-epoch behavior of PMS X-ray emission, \citet{Getman2024} also analyzed {\it Chandra} data from previous observations of the Orion Nebula region, collected in 2003, 2012, and 2016.

Briefly, in \citet{Getman2024}, standard {\it Chandra} data reduction tools from CIAO v4.15.2 \citep{Fruscione2006} were applied for data reprocessing, cleaning, and generating exposure maps. Point sources were identified using maximum likelihood image deconvolution with local point spread functions \citep{Broos2010}. X-ray point sources were then extracted and characterized using the ACIS Extract (AE) software package (version 5658, 2022-01-25) \citep{Broos2010,Broos2012}. By calibrating the 2023 {\it Chandra} X-ray light curves of PMS Orion members against the deeper 2003 {\it Chandra} Orion Ultradeep Project (COUP) observations \citep{Getman05} and employing the Bayesian Blocks algorithm \citep{Scargle98,Scargle2013}, over 150 Orion PMS stars exhibiting X-ray super-flares with energies $E_X>10^{34}$~erg were identified in the December 2023 observations. 

As described in \citet{Getman2024}, the X-ray count rates corresponding to the epoch-2023 Bayesian Blocks segments for each COUP star of interest were converted into intrinsic X-ray luminosities in the $(0.5-8)$~keV energy range ($L_X$). This conversion utilized AE-derived, instrument-independent, and point-spread-function (PSF)-corrected apparent X-ray photon fluxes ($F_{X,phot}$), measured in units of photons~cm$^{-2}$~s$^{-1}$, along with a single conversion factor relating $F_{X,phot}$ to $L_X$. 

The conversion factor was derived from the time-averaged X-ray luminosity values obtained from the epoch-2003 COUP dataset \citep{Getman05}. These $L_X$ values were calculated by \citet{Getman05} using {\it XSPEC} \citep{Arnaud1996} to fit the COUP spectra with a two-temperature, optically thin thermal plasma model ({ \it mekal$+$mekal }), subject to photoelectric absorption ({\it wabs}). The coronal metallicity in this modeling was fixed at $Z = 0.3$, consistent with the sub-solar coronal metallicities typically observed in young stars \citep{Gudel2007, Maggio2007, Schulz2024}. The fitted epoch-2003 COUP spectra and corresponding X-ray parameters are available in the electronic version of Figure Set 12 from \citet{Getman05}.

For the four HET-HPF stars of interest in the current paper (COUP 881, 1333, 1424, and 1463; see details below), the inferred epoch-2003 plasma temperatures for the soft component are around $kT_1 \sim 0.8-0.9$~keV. These temperatures may correspond to compact coronal loop structures covering the surfaces of active stars \citep{Preibisch05}. The hard plasma component, with temperatures $kT_2$ varying between 2 and 3~keV, likely originates from larger structures associated with more powerful flaring events \citep{Getman2008a}. The X-ray column densities, $\log(N_H)$, range between $20$ and $21.4$~cm$^{-2}$, indicating relatively low stellar absorption \citep{Feigelson05}.

Among $>150$ Orion PMS super-flaring stars identified in the epoch-2023 {\it Chandra} observations, 80 produced powerful flares with energies between $10^{35}$ and $10^{36}$~erg, while eight experienced mega-flares exceeding $E_X>10^{36}$~erg. 

The observed frequency of 88 flares with $E_X > 10^{35}$~erg in the Orion Nebula, within the {\it Chandra}-ACIS-I field over a total exposure time of 67~ksec, aligns with the predicted average flare rate of approximately 20 such flares per PMS star per year. This prediction is supported by earlier COUP epoch-2003 data and multi-epoch PMS X-ray flare studies across various Galactic star-forming regions \citep{Getman2021}. Notably, while PMS stars exhibit minor long-term baseline X-ray variations likely due to magnetic dynamo cycles \citep{Getman2024}, the consistency of large X-ray flare rates suggests these variations do not influence flare frequencies. A similar conclusion is drawn from X-ray studies of the older PMS system AB Dor \citep{Singh2024}.

Among the 88 Orion PMS stars exhibiting powerful X-ray flares with $E_X > 10^{35}$~erg, four were selected for follow-up HET-HPF observations based on the following criteria: K- or M-type stars, for which the planned magnetic intensification technique to measure magnetic field strengths, utilizing Ti lines at $960 - 980$~nm \citep{Kochukhov2021}, offers sufficient sensitivity; NIR-bright stars ($J<11.5$~mag) to ensure a reasonable signal-to-noise ratio (SNR) in the HET spectrum; and relatively slow rotators with mild or no accretion activity, allowing for spectral lines that are not significantly broadened by rotation or accretion. The four X-ray flaring stars (COUP881, 1333, 1424, and 1463), which met these criteria, were promptly observed by HET-HPF following their large X-ray flares.

\begin{deluxetable*}{ccccccccccc}
\tabletypesize{\normalsize}
\tablecaption{Basic stellar and {\it Chandra} X-ray properties of HET-HPF Targets  \label{tab:chandra_het_targets}}
\tablewidth{0pt}
\tablehead{
\colhead{Src.} & \colhead{SpT} & \colhead{$A_V$} & \colhead{$M$} & \colhead{$R$} & \colhead{$\alpha_{IRAC}$} & \colhead{$P_{rot}$} & \colhead{$v_{rot} sin(i)$} & \colhead{$\log(L_{X,char})$} & \colhead{$\log(L_{X,fl,pk})$} & \colhead{$\log(E_{X,fl})$}  \\
\colhead{} & \colhead{} & \colhead{(mag)} & \colhead{(M$_{\odot}$)} & \colhead{(R$_{\odot}$)} & \colhead{} & \colhead{(day)} & \colhead{(km~s$^{-1}$)} & \colhead{(erg~s$^{-1}$)} & \colhead{(erg~s$^{-1}$)} & \colhead{(erg)} \\
\colhead{(1)} & \colhead{(2)} & \colhead{(3)} & \colhead{(4)} & \colhead{(5)} & \colhead{(6)} & \colhead{(7)} & \colhead{(8)} & \colhead{(9)} & \colhead{(10)} & \colhead{(11)}
}
\startdata
COUP~881 & M1 & 0.0 & 0.6 & 1.7 & \nodata & \nodata & 9.1 & $30.49 \pm 0.043$& $31.69 \pm 0.055$ & $35.56 \pm 0.028$ \\
COUP~1333 & M0.5 & 1.1 & 0.8 & 2.2 & -0.9 & 9.0 & 13.8 & $30.28 \pm 0.066 $ & $32.15 \pm 0.010$ & $36.27 \pm 0.008$  \\
COUP~1424 &  M1 & 2.1 & 0.9 & 1.9 & -2.8 & 10.6 & 12.2 & $30.67 \pm 0.050$ & $31.60 \pm 0.015$ & $36.11 \pm 0.011$ \\
COUP~1463 &  M1 & 1.5 & 0.9 & 2.5 & -1.5 & \nodata & 13.1 & $30.29 \pm 0.032$ & $31.22 \pm 0.024$ & $35.35 \pm 0.022$ \\
\enddata 
\tablecomments{Four stars exhibiting prominent X-ray flares in December 2023 and observed with HET-HPF. Column 1: Simbad source name. Columns 2-8: Stellar properties including spectral type, visual extinction, stellar mass and radius, apparent slope in the spectral energy distribution from IR-band {\it Spitzer}-IRAC data, stellar rotation period, and projected velocity \citep{Getman2024, Megeath2012, Davies2014,DaRio2016,Kounkel2019}. Note that the $v_{\mathrm{rot}} \sin i$ stellar rotation velocities inferred from our HPF analysis (\S~\ref{sec:ZB_mesurements}) are systematically lower than those obtained from the relatively lower-resolution APOGEE spectra by \citet{DaRio2016, Kounkel2019}. The $\alpha_{IRAC}$ values of $<-1.9$ for COUP~1333 and COUP~1463 suggest that these stars possess a protoplanetary disk. Columns 9-11: X-ray luminosities (including 1-$\sigma$ uncertainties) for characteristic and flare peak levels, as well as X-ray flare energy \citep{Getman2024}. In case of COUP~881, the flare's peak X-ray luminosity and energy values are given for the first flare observed during the second {\it Chandra} observation (Figure~\ref{fig:chandra_lc}). 
}
\end{deluxetable*}

\begin{deluxetable*}{cccccccc}
\tabletypesize{\small}
\tablecaption{Magnetic Field Strengths Inferred from HPF Data  \label{tab:mag_field_strengths}}
\tablewidth{0pt}
\tablehead{
\colhead{Src.} & \colhead{Visits} & \colhead{Dates of First \& Last Visits} & \colhead{$\delta t$} & \colhead{\bs} & \colhead{$f_0$} & \colhead{$f_2$} & \colhead{$f_4$} \\
\colhead{} & \colhead{} & \colhead{} & \colhead{(days)} & \colhead{(kG)} & \colhead{} & \colhead{} & \colhead{} \\
\colhead{(1)} & \colhead{(2)} & \colhead{(3)} & \colhead{(4)} & \colhead{(5)} & \colhead{(6)} & \colhead{(7)} & \colhead{(8)}  
}
\startdata
COUP~881 &  3& $ 2023-12-25 T05:58:00.47 T06:30:31.28  $ & 2.67 & $  2.46^{+0.12}_{-0.13} $ & $  0.06^{+0.06}_{-0.04} $ & $  0.63^{+0.07}_{-0.08} $ & $  0.30^{+0.05}_{-0.05}$\\
COUP~881 & 3& $ 2023-12-31 T05:33:05.91 T06:05:36.71  $ & 8.65 & $  2.39^{+0.09}_{-0.10} $ & $  0.05^{+0.04}_{-0.03} $ & $  0.70^{+0.06}_{-0.06} $ & $  0.25^{+0.04}_{-0.04}$\\
COUP~1333 & 3& $ 2023-12-24 T05:57:37.89 T06:30:08.66 $ & 1.72 & $  2.62^{+0.20}_{-0.20} $ & $  0.28^{+0.08}_{-0.10} $ & $  0.13^{+0.15}_{-0.09} $ & $  0.58^{+0.06}_{-0.07}$\\
COUP~1333 & 3& $ 2023-12-29 T05:37:08.03 T06:09:38.83 $ & 6.71 & $  2.79^{+0.20}_{-0.21} $ & $  0.21^{+0.09}_{-0.11} $ & $  0.18^{+0.16}_{-0.12} $ & $  0.61^{+0.06}_{-0.08}$\\
COUP~1424 & 2& $ 2023-12-18 T06:25:56.05 T06:43:10.09 $ & -3.06& $  2.97^{+0.10}_{-0.11} $ & $  0.02^{+0.03}_{-0.02} $ & $  0.46^{+0.05}_{-0.05} $ & $  0.51^{+0.04}_{-0.05}$\\
COUP~1463 & 3& $ 2023-12-23 T06:06:40.29 T06:39:11.11 $ & 0.73 & $  2.75^{+0.13}_{-0.14} $ & $  0.10^{+0.08}_{-0.06} $ & $  0.42^{+0.10}_{-0.11} $ & $  0.48^{+0.05}_{-0.05}$\\
COUP~1463 & 3& $ 2023-12-28 T05:48:39.12 T06:21:09.91$ & 5.71& $  2.67^{+0.12}_{-0.13} $ & $  0.09^{+0.08}_{-0.06} $ & $  0.47^{+0.10}_{-0.12} $ & $  0.44^{+0.05}_{-0.05}$\\
COUP~1463 & 3& $ 2024-01-03 T05:17:10.24 T05:49:41.06$ & 11.69& $  2.83^{+0.10}_{-0.10} $ & $  0.04^{+0.05}_{-0.03} $ & $  0.50^{+0.06}_{-0.07} $ & $  0.46^{+0.04}_{-0.05}$\\ 
\enddata
\tablecomments{Column 1: Simbad source name. Column 2: Number of HPF observation visits. Column 3: Start dates (UT) of the first and last HPF visits per night of observation. Column 4: Time difference (in days) between the latest flare event's Bayesian Block segment (depicted in red in Figure~\ref{fig:chandra_lc}) and the nearest HPF visit. The negative value for COUP 1424 indicates that an HPF visit occurred during the decay phase of the large X-ray flare. Column 5: Inferred average surface magnetic field strength with 95\% confidence interval. Columns 6-8: Inferred surface filling factors for magnetic field components with $B=0$, 2, and 4~kG.
}
\end{deluxetable*}

Table~\ref{tab:chandra_het_targets} provides a summary of the stellar properties, including the X-ray luminosities at characteristic (baseline or quiescent) levels ($L_{X,char}$) and during flare peaks ($L_{X,fl,pk}$), as well as the X-ray flare energies ($E_X$). Figure~\ref{fig:chandra_lc} presents the December 2023 {\it Chandra} light curves, plotted in units of intrinsic X-ray luminosity for the $0.5-8$~keV band, for the four HET-HPF stars. 

All four stars are NIR-bright early M-type stars with J magnitudes ranging from 10.8 to 11.5 \citep{Getman05} and masses between 0.6 and 0.9~M$_{\odot}$. They exhibit relatively slow rotation, with projected rotational velocities between 9 and 14~km~s$^{-1}$. Notably, two of these stars (COUP 1333 and 1463) are surrounded by protoplanetary disks but have low accretion rates, as noted by \citet{Getman05}. All four stars show minimal dust extinction, with $A_V$ values of no more than 2 magnitudes, suggesting they are likely members of the ONC cluster \citep{Feigelson05}. Among them, COUP~881 and COUP~1463 displayed exceptionally strong X-ray super-flares ($E_X > 2 \times 10^{35}$~erg), while COUP~1333 and COUP~1424 produced even more powerful mega-flares ($E_X > 10^{36}$~erg).

For the relatively short X-ray flare from COUP~881, for which the decay phase was nearly fully captured, the inferred December 2023 values for $L_{X,fl,pk}$ and $E_X$ are already close to the true flare values. For the mega-flares ($E_X > 10^{36}$~erg) observed in COUP~1424 and COUP~1333, the inferred $L_{X,fl,pk}$ and $E_X$ values are also expected to be reasonably close to the true values. This is supported by the fact that the occurrence rate of even more powerful PMS flares (e.g., $E_X > 3 \times 10^{36}$~erg) is low, approximately 0.1 flares per star per year for stars with mass less than 1M$_{\odot}$ \citep{Getman2021}, which translates to less than one such flare per December 2023 {\it Chandra} observations of the Orion Nebula region. However, for the moderately powerful super-flare ($E_X = 2 \times 10^{35}$~erg) from COUP~1463, the $L_{X,fl,pk}$ and $E_X$ values could indeed be lower limits of the true values.

Based on X-ray observations of some stellar flares, flare-induced coronal abundance changes may occur \citep[][and references therein]{Gudel2009}. However, no evidence 
for unusual abundances during large PMS flares were found by \citet{Maggio2007}. Specifically, Maggio et al. conducted a detailed investigation of coronal abundances in nearly 100 low-absorption COUP PMS stars using statistically rich, {\it Chandra} X-ray COUP epoch-2003 spectra from \citet{Getman05}. They found that the time-averaged abundances of these PMS stars, {\it normalized to solar photospheric values}, displayed a systematic pattern: the abundances of certain high-FIP (First Ionization Potential) elements (e.g., S, O, Ar, and Ne) were consistently higher than those of low-FIP elements (e.g., Mg, Fe, and Si). Furthermore, their Figure 10 demonstrates that the coronal abundances of Fe (representing low-FIP elements) and Ne (representing high-FIP elements) in COUP PMS stars undergoing very large X-ray flares are indistinguishable from those in COUP PMS stars without such flares.

The {\it Chandra} spectra of our two flaring stars, COUP~1424 and COUP~1463, were previously analyzed by \citet{Maggio2007} based on the epoch-2003 COUP observations. For both stars, the abundances inferred from their epoch-2003 spectra were sub-solar for nearly all elements ($\leq 0.4$), except for Ne and Ni, which is consistent with the choice of $Z=0.3$ in the simpler spectral fitting approach used by  \citet{Getman05}. Since the epoch-2023 {\it Chandra} spectrum of COUP~1424 is dominated by mega-flare emission (Figure~\ref{fig:chandra_lc}), we tested the hypothesis that coronal abundances might increase during this flare.

The epoch-2023 spectrum of COUP 1424, with over 3000 X-ray counts, was modeled using a series of trial cases with two-temperature plasma {\it mekal} or {\it apec} or {\it vapec} components, subject to {\it wabs}-based absorption. In all cases, the hydrogen column density and the lower-temperature component ($kT_1$) were fixed to the time-averaged values reported in \citet{Getman05}. The best spectral fits were achieved with either {\it wabs(apec+apec)} (or {\it wabs(mekal+mekal)} ) models assuming a global metallicity of $Z = 0.3$ or {\it wabs(vapec+vapec)} models using individual elemental abundances as determined by \cite{Maggio2007}, yielding reduced $\chi^2$ values around 1.1-1.2. In contrast, trial fits with metallicity fixed to systematically higher values ($Z =$~0.6, 1.0, and 1.5) produced progressively worse fits, with reduced $\chi^2$ increasing to 1.3, 1.5, and 1.8, respectively. These results show no indication of increased coronal abundances during the mega-flare of COUP 1424. 

It is important to emphasize that the time-averaged X-ray-derived coronal abundances are already consistent with optically determined photospheric abundances for at least Si, O, and Ne in PMS stars \citep{Maggio2007}. Therefore, flare-induced chromospheric evaporation is unlikely to significantly alter coronal abundances. More comprehensive PMS abundance studies, involving a comparative analysis of stacked {\it Chandra} spectra of numerous Orion Nebula PMS stars during large X-ray flare and quiescent states, fall outside the scope of this paper and will be addressed in future work.

Moreover, since the plasma in large X-ray flares is extremely hot, X-ray emission is dominated by the continuum component, rendering the choice of $Z$ negligible for determining the best-fit plasma temperature and X-ray luminosity.

\subsection{HET-HPF observations} \label{sec:hpf_observations}
The Habitable Zone Planet Finder is an ultra-stabilized, fiber-fed, near-infrared spectrograph \citep[HPF;][]{Mahadevan2012,Mahadevan2014} installed on the 11-m Hobby-Eberly Telescope at McDonald Observatory, Texas, USA \citep[HET;][]{Ramsey1998,Hill2021}. HPF operates over a wavelength range of 8080 to 12,780 \AA~ at a spectral resolution of R $\approx$ 55,000. In addition to the science fiber, which collects the stellar spectrum, the HPF is equipped with two other fibers for simultaneous calibration and monitoring of the sky background spectrum. Since HPF is a stabilized radial velocity spectrograph, simultaneous Laser Frequency Comb (LFC) calibration was not used during our observations. The HET is a queue-scheduled telescope \citep{Shetrone2007}. Within a day of detecting the X-ray flares, the four PMS stars exhibiting flares were added to the observation queue. Due to the restricted viewing altitude of the HET ($55 \pm 8$ degrees), only one target in the Orion declination could be observed per night. Two observations per target were planned. The first observation was made as soon as weather conditions permitted. The second observation was scheduled after half of the star's rotation period had elapsed since the first observation (Table~\ref{tab:mag_field_strengths}). This was done to avoid the possibility of the flaring region being hidden behind the visible disk during one of the observation epochs.

For COUP 881, 1333, and 1463, each star observation had a total exposure time of 2700 seconds, divided into three visits of 900 seconds each (Table~\ref{tab:mag_field_strengths}). The only observation of COUP 1424 included two visits. Data were reduced using the standard HPF data reduction pipeline. In brief, the up-the-ramp exposures were bias-corrected, cleaned of cosmic rays, and gain-corrected before generating the photoelectron counts-per-second map on the detector \citep{Ninan2018}. The stellar spectrum was extracted using a slightly modified version of the flat-relative optimal extraction algorithm \citep{Zechmeister2014}. The simultaneously observed sky fiber spectrum was scaled, interpolated, and subtracted from the stellar spectrum. Wavelength calibration was achieved by drift-correcting the absolute wavelength solution, which was determined using the HPF LFC \citep{Metcalf2019}, to the observation epoch of the spectrum \citep{Stefansson2020}. Telluric absorption lines in the stellar spectrum were corrected using an HPF PSF-convolved telluric grid model (Krolikowski et al., in preparation).

\section{Measurements of magnetic field strengths} \label{sec:ZB_mesurements}
In this study we used detailed spectrum synthesis calculations, including the effects of Zeeman splitting and polarized radiative transfer, to measure average magnetic field strengths in the four COUP targets. Based on a series of test calculations for the entire HPF wavelength range, we have established that a set of 10 Ti~{\sc i} lines\footnote{The Ti~{\sc i} $\lambda$ 9783.31 and 9783.59~\AA\ are blended into a single feature, so the analysis presented here is based on 9 spectral intervals.} with central wavelengths 9638--9788~\AA\ provides the best magnetic diagnostic given the early-M spectral types of our targets and the S/N variation with wavelength in the observed spectra. These neutral Ti lines originate from the same multiplet and thus have well-established relative strengths while exhibiting a diverse magnetic field response. Previous studies of magnetic fields in low-mass stars \citep[e.g.][]{Kochukhov2017,Shulyak2017,Shulyak2019,Reiners2022} demonstrated that a spectrum synthesis analysis based on these lines can disentangle magnetic field from various non-magnetic effects on spectral lines, retrieving information on the surface field strength for both slow rotators (utilizing Zeeman broadening) and for fast-rotating targets (taking advantage of the differential magnetic intensification).

For our analysis, the plane-parallel stellar model atmospheres from the {\sc MARCS} \citep{Gustafsson2008} solar-metallicity grid were used, with the model atmosphere parameters, $T_{\rm eff}$ and $\log g$, adopted from \citet{DaRio2016}. The line list was retrieved from {\sc VALD} data base \citep{Ryabchikova2015} and the spectrum synthesis calculations were carried with the {\sc Synmast} code \citep{Kochukhov2007,Kochukhov2010} assuming a uniform radial magnetic field. A discussion and justification of this commonly assumed magnetic field orientation can be found in e.g. \citet{Kochukhov2021}. We employed a multi-component magnetic field model \citep[e.g.][]{Johns-Krull2007,Reiners2022} to approximate a distribution of the magnetic field strengths at the stellar surface. Given the spectral resolution of HPF and small to moderate rotational broadening of the four COUP targets, it is appropriate to adopt a step of 2~kG for sampling the field strength distribution. The resulting composite synthetic model contains spectral contributions of 0~kG, 2~kG, 4~kG, etc. magnetic components with the corresponding filling factors $f_0$, $f_2$, $f_4$, etc. In the inference, the field strengths are fixed whereas the filling factors are allowed to vary. The mean magnetic field strength is then determined as $\langle B \rangle=2 f_2 + 4 f_4 + \dots$ from a set of filling factors adjusted to fit observations.

Several additional parameters have to be adjusted in the magnetic inference in addition to fitting the magnetic filling factors. This includes a radial velocity shift, non-magnetic line broadening, and line-strength parameter. The non-magnetic broadening was treated by convolving synthetic spectra with a Gaussian profile with a variable width for the narrow-line targets COUP~881 and COUP~1424. For the two remaining stars, COUP~1333 and COUP~1463, showing a non-negligible rotational broadening we used a fixed Gaussian profile to represent the instrumental broadening and a variable $v_{\rm rot}\sin i$ to account for the rotational broadening.

The HPF-based rotational broadening values are systematically smaller than those listed in Table~\ref{tab:chandra_het_targets}, which were obtained from relatively low-resolution ($R = 22,500$, FWHM $= 13.3$~km/s) APOGEE spectra by \citet{DaRio2016} and \citet{Kounkel2019}. With the significantly higher resolution of HPF data and a more sophisticated spectral analysis employed in this work --- one that incorporates magnetic broadening effects --- we have improved the determination of $v_{\rm rot}\sin i$. The HPF-based $v_{\rm rot}\sin i$ values for COUP~1333 and COUP~1463 are $6.5 \pm 0.8$~km/s and $5.0 \pm 1.2$~km/s, respectively, while the $v_{\rm rot}\sin i$ upper limits for COUP~881 and COUP~1424 are $1{-}2$~km/s.

Regarding the line strength parameter, we found that adjusting the veiling factor \citep[e.g.][]{Lavail2019} with the assumed solar Ti abundance yields satisfactory fit for all targets except COUP~1424, which exhibits too strong Ti lines even with zero veiling. For that target we varied the Ti abundance to fit the observations.

\begin{figure*}
    \centering
     \includegraphics[width=0.95\textwidth]{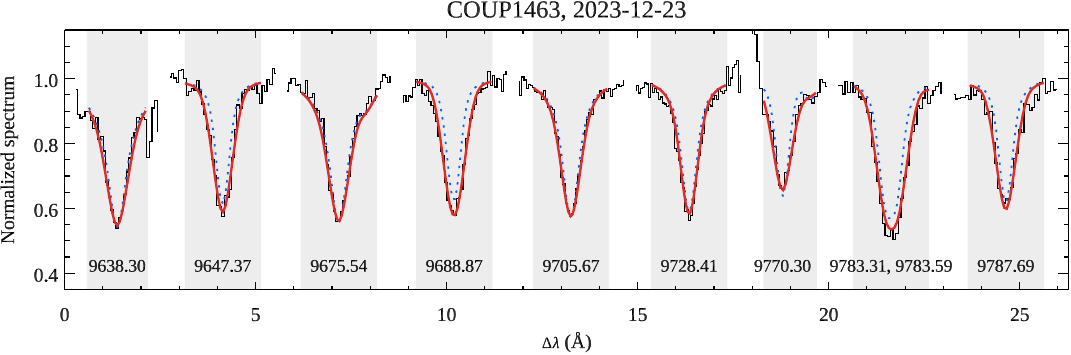}
    \caption{Illustration of the magnetic field strength determination for COUP~1463 observed on 2023-12-23. The black histogram shows the observed spectrum corrected for telluric absorption. The red line corresponds to the best-fitting magnetic model spectrum, which includes $B=0$, 2, and 4~kG components as documented in Table~\ref{tab:mag_field_strengths}. For comparison, the blue dotted line shows calculation with the same stellar parameters but with the magnetic field turned off. The central wavelengths of 9 Ti~{\sc i} spectral features employed for the magnetic field measurement are indicated below each line.} \label{fig:mag_fit}
\end{figure*}

We ran the Markov chain Monte Carlo (MCMC) calculations using the IDL {\sc SoBat} package by \citet{Anfinogentov2021} to sample all model parameters, derive the best-fitting parameters, and calculate realistic uncertainties. Multiple observations of different targets were analyzed independently. For each observed spectrum, we carried out multiple MCMC runs with different maximum magnetic field strengths, obtaining best-fitting parameters from the median of posterior distributions. These fits were compared using the Bayesian information criterion \citep[BIC,][]{Sharma2017}, which allowed us to objectively determine the strongest-field magnetic component justified by observations. We found that for all targets the BIC analysis favoured a superposition of 2 and 4~kG magnetic components alongside a non-magnetic spectral contribution.

The average magnetic field strengths and individual filling factors are reported in Table~\ref{tab:mag_field_strengths} for all 8 observations of four COUP targets. We measured field strengths between 2.4 and 3~kG with a typical 1-$\sigma$ uncertainty of 0.15~kG. An example of the fit to the HPF observation of COUP~1463 obtained on 2023-12-23 is shown in Figure~\ref{fig:mag_fit}. This plot presents both magnetic and non-magnetic calculations for the same set of nuisance parameters, illustrating different sensitivity of the studied Ti~{\sc i} lines to the Zeeman effect.

\section{RESULTS} \label{sec:results}

\subsection{No indication of abnormally strong magnetic fields} \label{sec:typical_B_fields}
The observation dates for HET-HPF listed in Table~\ref{tab:mag_field_strengths} show that HPF began sampling magnetic field strengths during the decay phase of the large X-ray flare in COUP~1424. For COUP~1463, COUP~1333, and COUP~881 magnetic field recordings began approximately 0.7, 1.7,  and 2.7 days after the last known flare segments of their respective large X-ray flares.

Table~\ref{tab:mag_field_strengths} lists multiple HET-HPF observations for three of the four stars, reflecting our efforts to sample the magnetic field at different stellar rotational phases. However, none of the stars (COUP 881, 1333, and 1463) with multiple observations display significant variation in magnetic field strength across phases. This absence of rotational modulation in surface magnetic field strength is supported by our quantitative finding of extensive surface coverage by magnetic active regions, with surface filling factors $f_2 + f_4$ reaching or exceeding 80-90\% of the stellar surface area.

For further analyses, for COUP~881, 1333, and 1463, the values of their magnetic field strength and filling factors are averaged across multiple HET-HPF observations.

\begin{figure*}
    \centering
     \includegraphics[width=0.95\textwidth]{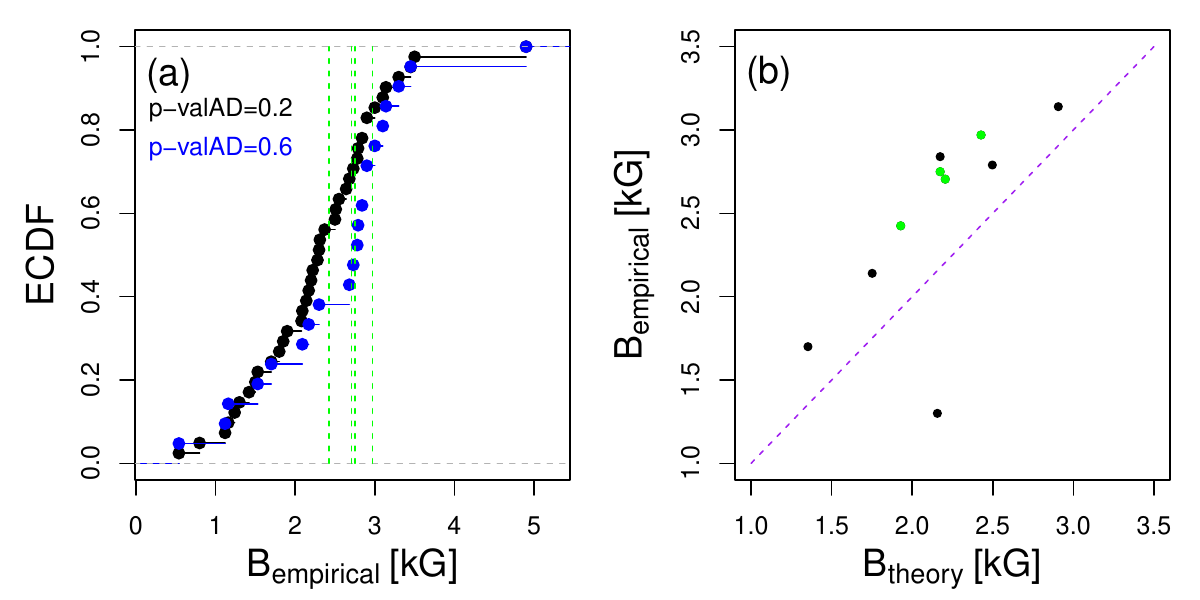}
    \caption{Comparison of the HPF-based surface magnetic field strengths inferred for the four flaring Orion PMS stars with: (a) the PMS magnetic field strength compilation from \citet{Sokal2020} and (b) predictions from convection-driven dynamo theory \citep{Christensen2009, Reiners2022}. In both panels, the four flaring Orion PMS stars are highlighted in green. Panel (a) displays the full PMS sample from \citet{Sokal2020} as a black Empirical Cumulative Distribution Function (ECDF) curve, and its M-type stellar sub-sample is represented in blue. Anderson-Darling (AD) p-values for the comparisons between the \citet{Sokal2020} (sub)-samples and the ECDF of our four flaring stars are listed in the legend. Panel (b) additionally shows six more COUP stars from \citet{Sokal2020} (black), with their masses, bolometric luminosities, and radii estimated similarly to those of our four flaring stars \citep{Getman22}. The unity line is depicted in purple.} \label{fig:B_comparison}
\end{figure*}

Figure~\ref{fig:B_comparison} presents a comparison of the average surface magnetic field strengths, derived from Zeeman broadening (ZB) measurements, for our four flaring Orion PMS stars with those from a compilation of over 40 fully convective PMS stars in \citet{Sokal2020}. Additionally, the figure contrasts these measurements with theoretical predictions for surface magnetic field strengths based on convection-driven dynamo models.

The high p-values obtained from the Anderson-Darling test \citep{AndersonDarling1952} shown in panel (a) suggest that the \bs\ measurements for the four flaring Orion PMS stars (green) are statistically consistent with the \bs\ distributions of both the full sample (black) and the M-type star sub-sample (blue) of primarily non-flaring PMS stars from \citet{Sokal2020}. 

Although magnetic filling factors are often not tabulated in the papers cited by \citet{Sokal2020}, the overall inferred high average magnetic field strengths in PMS stars imply correspondingly high magnetic filling factors ($>60-80$\%) for these stars. These values are significantly higher than the filling factors ($<5-20$\%) observed in older solar-type stars \citep{Kochukhov2020}, indicating that the convection-driven dynamo in PMS stars is more effective in generating large active regions compared to the tachocline dynamo in older solar-type stars.

For stars with rapid rotation and convection-driven dynamos, such as early PMS stars or old late-M dwarfs, the available kinetic energy in their convective motions may determine the strength of their surface magnetic fields \citep{Christensen2009, Reiners2009, Reiners2010, Reiners2022}. These studies offer a straightforward analytical expression for estimating the average surface magnetic field strength. In this formula, the magnetic pressure ($\propto B^2$) is influenced by the stellar density ($\rho^{1/3}$, where $\rho \sim M/R^3$) and the bolometric flux ($q^{2/3}$, where $q \sim L/R^2$). When calibrated to the empirical values of magnetic field strengths for different stars and planets (in units of Gauss), the equation becomes: \begin{equation} B_{kin} = 4800 (M L^2/R^7)^{1/6} \label{eqn:Bkin} \end{equation} where $M$, $L$, and $R$ represent the stellar mass, bolometric luminosity, and radius, all expressed in solar units.

We applied the above formula to our four flaring Orion PMS stars, as well as to a subset of Orion PMS stars from the \citet{Sokal2020} compilation, which are part of the 2003 COUP {\it Chandra} studies \citep{Getman05}. The masses, bolometric luminosities, and radii for these stars were estimated using the same methods as for our four flaring HET-HPF stars \citep{Getman22}. In panel (b), 9 out of the 10 stars show empirical ZB \bs\ measurements that align with the theoretical predictions derived from the formula, assuming a small systematic calibration adjustment by a factor of 1.2 between the theoretical model and the observational data.

To summarize, results presented in Figure~\ref{fig:B_comparison} show that the average surface magnetic field strengths measured in our four Orion PMS stars during or shortly after their large X-ray flares are consistent with values typical of non-flaring PMS stars. These empirical measurements of \bs\ also align with the basic predictions of convection-dynamo theory. Consequently, these results effectively rule out our hypothesis of unusually strong magnetic fields occurring during or after large PMS X-ray flares.

\begin{figure*}
    \centering
     \includegraphics[width=0.98\textwidth]{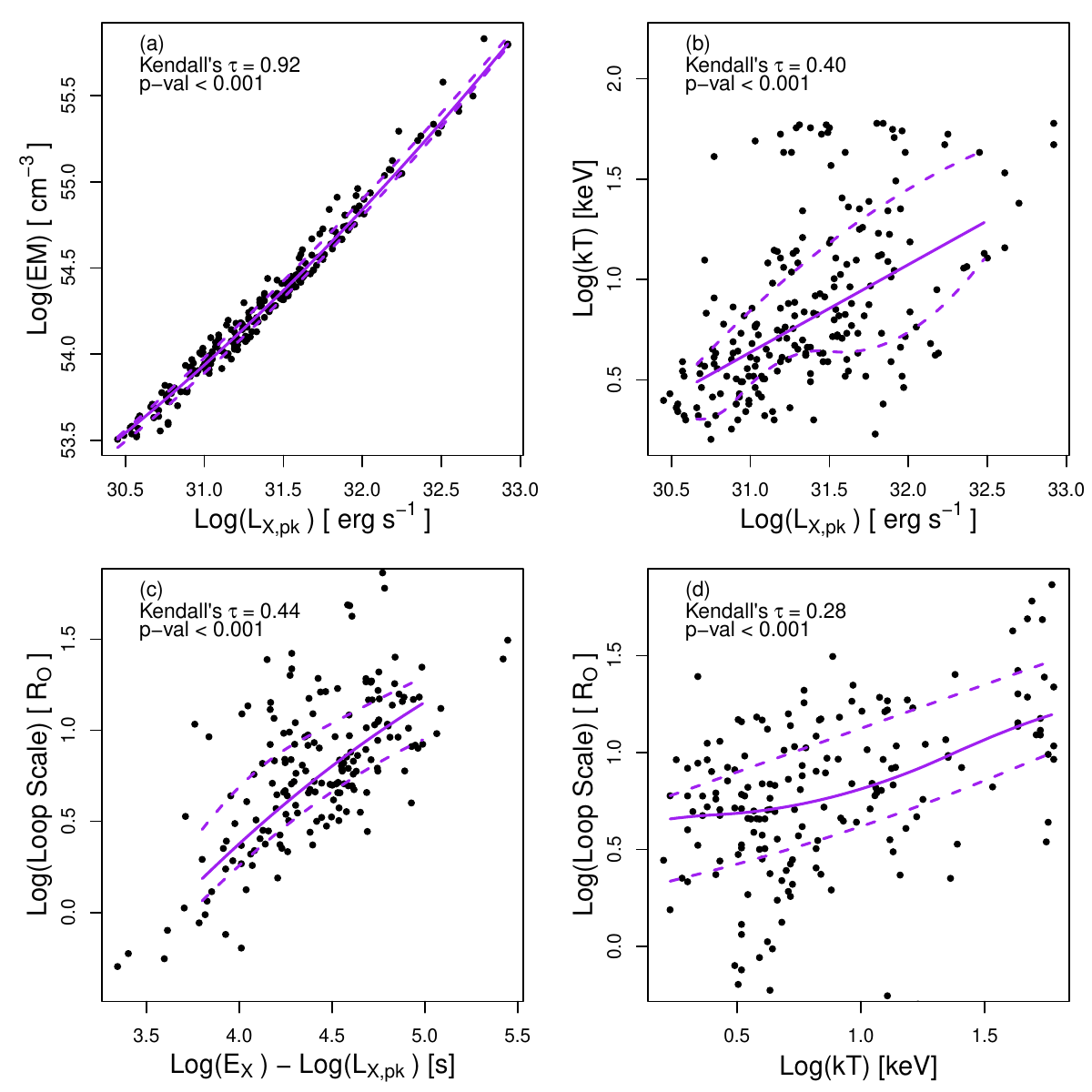}
    \caption{Depiction of the guiding COUP flare trends based on the COUP flare sample and flare parameters from \citet{Getman08a}. (a,b) Emission measure and plasma temperature as functions of the X-ray flare peak luminosity. (c,d) Flare loop scale as a function of approximate flare duration and plasma temperature. The flare duration is estimated as the $\log$ difference between the flare energy and flare peak luminosity. The purple curves represent spline fits to the 25\%, 50\%, and 75\% quartiles of the distributions, created using R's Constrained B-Splines cobs function \citep{Ng2007}. The corresponding legends display the $\tau$ coefficients and p-values from Kendall's $\tau$ test, which indicate a strong positive correlation in the left panel ($\tau = 0.92$) and moderate positive correlations in the other three panels ($\tau \geq 0.28$), all of which are highly statistically significant.} \label{fig:COUP_calibration_curves}
\end{figure*}

\begin{figure*}
    \centering
     \includegraphics[width=0.95\textwidth]{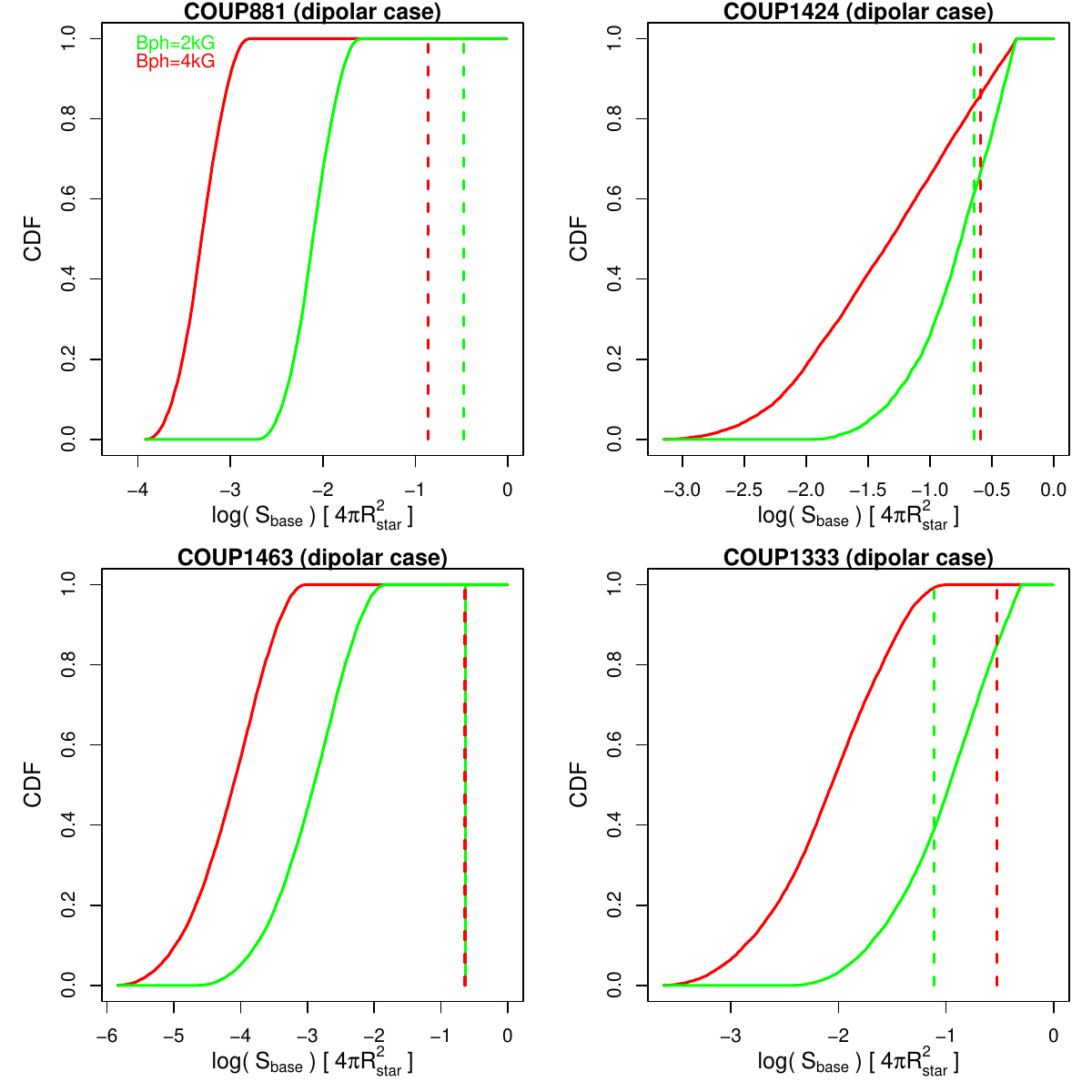}
    \caption{The results from the simulations are shown as cumulative distribution functions (CDFs) of flaring loop cross-sections in units of stellar surface area for a dipolar magnetic loop topology (solid curves). These are presented for two values of photospheric magnetic field strength --- 2~kG (green) and 4~kG (red). The vertical dashed lines indicate half of the empirically derived ZB-based magnetic filling factors, also in units of stellar surface area.} \label{fig:simulation_results_dipolar_case}
\end{figure*}

\begin{figure*}
    \centering
     \includegraphics[width=0.95\textwidth]{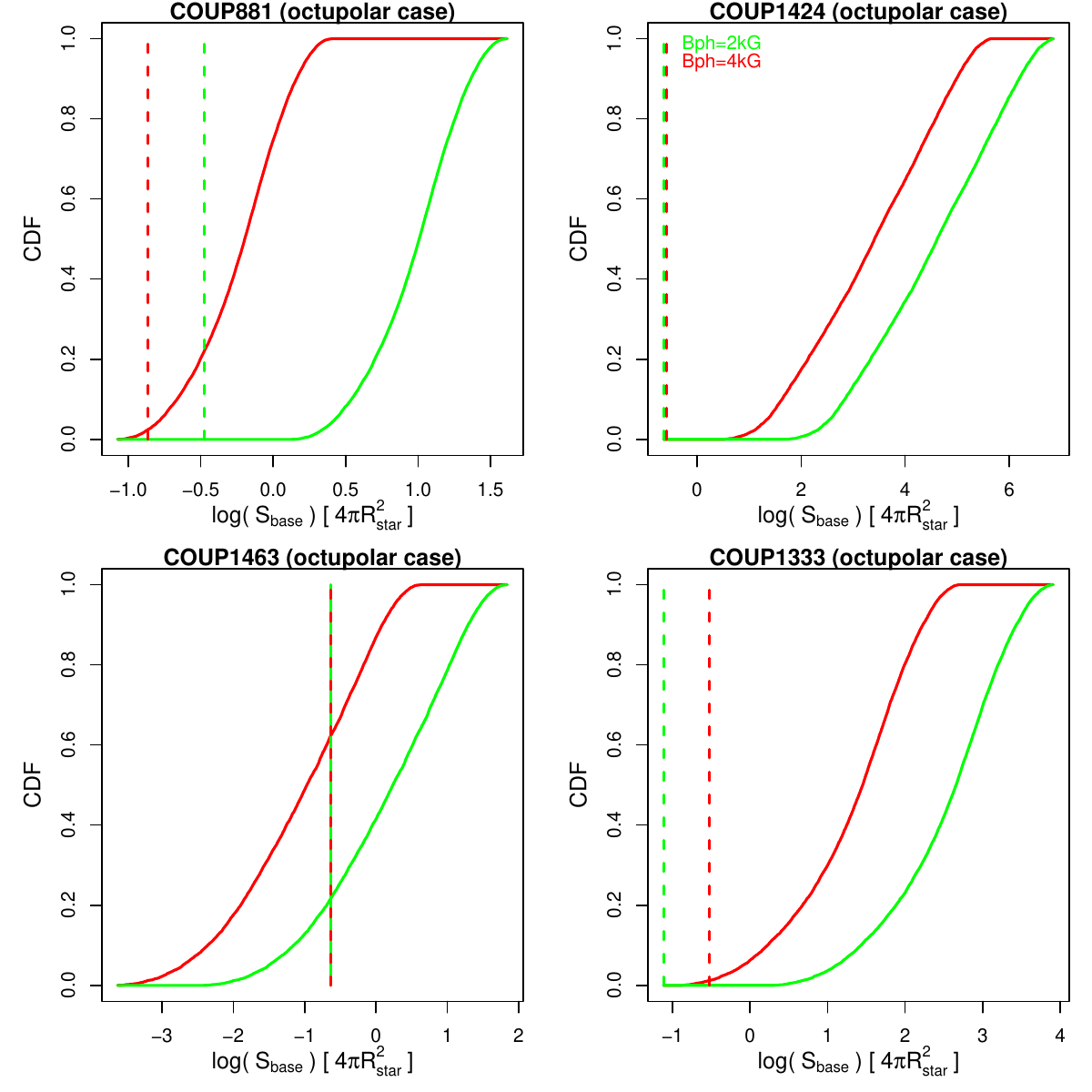}
    \caption{Same as  Figure~\ref{fig:simulation_results_dipolar_case}, but for an octupolar magnetic loop topology.} \label{fig:simulation_results_octupolar_case}
\end{figure*}

\begin{deluxetable*}{ccccccc}
\tabletypesize{\normalsize}
\tablecaption{Cross-Sections Of The Loops At Their Footpoints  \label{tab:mag_loop_crosssections}}
\tablewidth{0pt}
\tablehead{
\colhead{Src.} & \colhead{Topology} & \colhead{$B_{ {\text phot} }$} & \colhead{$S_{\text{HPF}}/2$} & \colhead{$S_{ \text{base},25\%}$} & \colhead{$S_{ \text{base},50\%}$} & \colhead{$S_{ \text{base},75\%}$} \\
\colhead{} & \colhead{} & \colhead{(kG)} & \colhead{$4 \pi R_{star}^2$} & \colhead{$4 \pi R_{star}^2$} & \colhead{$4 \pi R_{star}^2$} & \colhead{$4 \pi R_{star}^2$}\\
\colhead{(1)} & \colhead{(2)} & \colhead{(3)} & \colhead{(4)} & \colhead{(5)} & \colhead{(6)}  & \colhead{(7)} 
}
\startdata
COUP 881 & Dip & 2 & 0.334 & 0.0053 & 0.0078 & 0.0115 \\
COUP 881 & Dip & 4 & 0.137 & 0.0003 & 0.0005 & 0.0007 \\
COUP 1333 & Dip & 2 & 0.077 & 0.0483 & 0.1113 & 0.2262 \\
COUP 1333 & Dip & 4 & 0.297 & 0.0036 & 0.0089 & 0.0214 \\
COUP 1424 & Dip & 2 & 0.228 & 0.0948 & 0.1780 & 0.3067 \\
COUP 1424 & Dip & 4 & 0.257 & 0.0141 & 0.0489 & 0.1558 \\
COUP 1463 & Dip & 2 & 0.231 & 0.0005 & 0.0013 & 0.0031 \\
COUP 1463 & Dip & 4 & 0.229 & 2.96e-5 & 7.99e-5 & 0.0002 \\
\enddata 
\tablecomments{Column 1: SIMBAD source name. Column 2: Magnetic loop configuration: Dipolar (Dip). Column 3: Photospheric magnetic field strength. Column 4: Half of the magnetic surface area derived from HPF-based ZB analysis (indicated by dashed lines in Figure~\ref{fig:simulation_results_dipolar_case}). Columns 5-7: 25th, 50th, and 75th percentiles of the loop cross-section distributions (indicated as solid curves in Figure~\ref{fig:simulation_results_dipolar_case}) obtained from the simulations.}
\end{deluxetable*}

\subsection{Cross sections of flaring coronal loops} \label{sec:cross_sections_flaring_loops}

\subsubsection{Previously estimated cross sections} \label{sec:previous_loop_analyses}
First, we investigated the $L_X - \Phi$ positions of the previously analyzed 21 powerful PMS X-ray flares (shown as brown arrows in Figure~\ref{fig:pevtsov_plot}).

The loop scales for these 21 powerful PMS X-ray flares observed across various nearby Galactic star-forming regions were determined by \citet{Getman2021b} based on empirical X-ray measurements of flare duration ($\tau_{\mathrm{flare}}$), plasma temperature ($T$), and the slope on the temperature-density diagram ($\zeta$), utilizing the relation $\text{Loop Scale} = f(\tau_{\mathrm{flare}}, T, \zeta)$ derived from hydrodynamical simulations of solar/stellar coronal flaring loops \citep{Reale1997}. 

Subsequently, the temporal evolution of thermal conduction and radiation loss timescales for the flaring plasma were calculated as functions of plasma temperature, X-ray emission measure, loop scale, peak X-ray luminosity, and loop cross-section over a wide range of possible cross-section values. The combined cooling timescale was then compared with the empirical flare decay timescale to determine the most appropriate value for the loop cross-section. The loop was modeled as a cylinder with a constant cross-section ($S(z) = S_{\text{base}}$). 

Assuming magnetothermal pressure equilibrium along dipolar magnetic field lines, the underlying photospheric magnetic field strength $B_{\mathrm{phot}}$ was estimated (which was found to be close to or below 1~kG). The magnetic flux passing through the loop was inferred as $B_{\mathrm{phot}} \times S_{\mathrm{base}}$. The positions of these powerful PMS X-ray flares on the $L_X - \Phi$ diagram are shown in Figure~\ref{fig:pevtsov_plot} as brown arrows, indicating possible lower limits on magnetic fluxes due to the relatively low estimated $B_{\mathrm{phot}}$ (lower than the typical $B_{\mathrm{phot}}$ observed in PMS stars; see Figure~\ref{fig:B_comparison}).

The major limitation in \citet{Getman2021b}'s calculations lay in the inconsistent assumptions --- namely, modeling a giant flaring loop with a constant cross-section while assuming that the magnetic field strength followed a dipolar pattern, which contradicts the principle of magnetic flux conservation. This inconsistency, along with the use of real observed photospheric magnetic field strengths, has been corrected in our analysis described below.

\subsubsection{Cross sections of flaring loops in four HPF stars} \label{sec:new_loop_analyses}

To verify the placement of large X-ray PMS flares on the $L_X - \Phi$ diagram (Figure~\ref{fig:pevtsov_plot}) using our new empirical findings on flare peak X-ray luminosities, energies, and surface magnetic field strengths (Tables~\ref{tab:chandra_het_targets},\ref{tab:mag_field_strengths}), we begin by estimating the cross-sectional areas of the flaring coronal loops at their footpoints.

For this analysis, we make two key assumptions: the dominance of a giant PMS flaring magnetic coronal loop \citep{Favata2005,Getman2011,Reale2014} with either a dipolar or octupolar topology \citep{Donati2009}; and an equipartition magnetic field strength, where the magnetic pressure is comparable to the thermal pressure of the X-ray emitting plasma within the flaring coronal loop \citep{Favata2005}. The cross-sectional area of the flaring loop can then be estimated using a series of equations:

\begin{equation}
B_{\text{loop}}  \simeq \frac{B_{\text{phot}}}{\left( \frac{\text{LoopScale}}{R_{\star}} + 1 \right)^{\alpha}}
\label{eqn:Bloop}
\end{equation}

In this equation, $B_{\text{loop}}$ represents the magnetic field strength in the coronal loop at a height $\text{Loop Scale}$, which is determined by the underlying photospheric magnetic field strength $B_{\text{phot}}$ of a star with radius $R_{\star}$. The power-law index $\alpha$ corresponds to the magnetic field topology, where $\alpha = 3$ for a dipolar configuration and $\alpha = 5$ for an octupolar configuration.

Assuming equipartition, the electron density ($n_e$) of the X-ray emitting plasma within the loop is given by:

\begin{equation}
n_e = \frac{B_{\text{loop}}^2}{8 \pi K_b T}
\label{eqn:n_e}
\end{equation}

where $T$ is the plasma temperature and $K_b$ is the Boltzmann constant. The volume ($V$) of the X-ray emitting plasma is then:

\begin{equation}
V = \frac{\text{EM}}{n_e^2}
\label{eqn:V}
\end{equation}

where $\text{EM}$ is the X-ray emission measure. 

Finally, based on equation (\ref{eqn:Bloop}) and the principle of magnetic flux conservation along the coronal loop given as:

\begin{equation}
S(z) = S_{\text{base}} \times \frac{B_{\text{phot}}}{B(z)}
\label{eqn:flux_conservation}
\end{equation}

where $S_{\text{base}}$ is the cross-sectional area of the loop at its base, and solving the integral for the loop's volume:

\begin{equation} 
V = 2 \times \int_0^{\text{LoopScale}} S(z) dz 
\label{eqn:integral_for_volume}
\end{equation}

the cross-sectional area at the base of the loop can be expressed in terms of the loop's volume, $\text{LoopScale}$, and stellar radius as:

\begin{equation} 
S_{\text{base}} = \frac{0.5 \times (\alpha + 1) \times V \times R_{\star}^{\alpha}}{ ( \text{LoopScale} + R_{\star})^{\alpha+1} - R_{\star}^{\alpha+1} } 
\label{eqn:loop_scale}
\end{equation} 

To estimate $S_{\text{base}}$ using the above equations, we need the following parameters: photospheric magnetic field strength ($B_{\text{phot}}$), coronal loop scale ($\text{Loop Scale}$), peak flare X-ray plasma temperature ($T$), and emission measure ($\text{EM}$). The values for $B_{\text{phot}}$ are derived from HPF-based measurements (\S~\ref{sec:ZB_mesurements}), where we assume two cases: $B_{\text{phot}} = 2$~kG and $B_{\text{phot}} = 4$~kG.

The December 2023 Chandra data --- comprising six short observations separated by significant gaps (see Figure~\ref{fig:chandra_lc}) --- lack the rich X-ray photon statistics required for detailed time-resolved flare analysis and modeling \citep{Favata2005,Getman2011,Getman2021b}. Furthermore, information on the morphology and duration of the large X-ray flares detected in our four HET-HPF stars is limited. Instead, typical ranges for $\text{Loop Scale}$, $T$, and $\text{EM}$ for large PMS X-ray flares can be inferred from the COUP-based flare calibration curves. For this purpose, we reasonably assume that the December 2023 X-ray flares follow the trends observed in large COUP X-ray flares from the 2003 Chandra observations, as analyzed and modeled by \citet{Getman08a,Getman08b}.

Figure~\ref{fig:COUP_calibration_curves} shows that typical interquartile ranges (IQRs) for the loop length\footnote{Note that the \citet{Reale1997} model assumes that loop growth is unrestrained by gravity. This condition is always satisfied for early PMS flaring loops, even those as high as $10$--$20$~R$_{\star}$, because the pressure scale height always exceeds the loop height \citep{Getman2011}. Additionally, only a handful of points in Figure~\ref{fig:COUP_calibration_curves} correspond to loop scales exceeding 10 stellar radii.}, plasma temperature, and emission measure of the large X-ray flares in our four HPF stars can be estimated based on the measured values of their flare peak X-ray luminosity and flare energy, listed in Table~\ref{tab:chandra_het_targets}. 

Based on this information, we estimate the distributions of permissible loop cross-sections. Using hundreds of Monte Carlo simulations, values are randomly drawn from the interquartile ranges (IQRs) of loop scale, plasma temperature, and emission measure. The loop scale and plasma temperature values are additionally constrained to the 25$-$75 percentile range of COUP flares, as shown in Figure~\ref{fig:COUP_calibration_curves}d.  The formulas (\ref{eqn:Bloop}–\ref{eqn:loop_scale}) are then used to obtain distributions of loop cross-sections. These calculations account for two values of photospheric magnetic field strength (2~kG and 4~kG) and two coronal magnetic loop topologies (dipolar and octupolar). Assuming that giant PMS flaring loops have both footpoints anchored on the stellar surface \citep{Getman08b, Getman2021b, Getman2024}, loop cross-sections that exceed half of the total stellar surface area are excluded as astrophysically implausible.

Figures~\ref{fig:simulation_results_dipolar_case} and \ref{fig:simulation_results_octupolar_case} present the results of our simulations for dipolar and octupolar magnetic configurations, respectively. The relative positioning of the vertical dashed lines, based on HPF data, and the solid curves derived from the simulations delineates the permissible ranges for loop cross-sections. Specifically, values from the modeled distributions (the solid curves) that fall to the left of the observed values (the dashed lines) are considered astrophysically viable, as they allow the footpoints of flaring coronal loops to fit within the surface active regions, whose sizes are determined by HPF data.

For the X-ray flares in COUP~881, COUP~1424, and COUP~1333, no valid loop cross-section solutions are found for an octupolar topology, as nearly all simulated $S_{\text{base}}$ values exceed the sizes of the empirically determined HPF-based active regions, with many also surpassing the stellar surface areas (Figure~\ref{fig:simulation_results_octupolar_case}). For COUP~1463, fewer than 50\% of the simulated solutions are valid. Consequently, octupolar cases are excluded from further analyses.

Additionally, for the mega-flares in COUP~1424 and COUP~1333, the solutions with dipolar loops sustained by 4~kG active regions are more plausible than the ones supported by 2~kG active regions (Figure~\ref{fig:simulation_results_dipolar_case}).

For the powerful super-flares in COUP~881 and COUP~1463, all types of dipolar solutions --- whether with 2~kG or 4~kG active regions --- are possible. 

Table~\ref{tab:mag_loop_crosssections} presents the most plausible loop cross-section solutions derived from the dipolar case simulations. In summary, among the permissible dipolar solutions, the cross-sectional areas at the base of the loop ($S_{\text{base}}$) are systematically larger for the mega-flaring loops of COUP~1333 and COUP~1424 compared to the super-flaring loops of COUP~881 and COUP~1463. For the mega-flares, the interquartile ranges (IQRs) of loop cross-sections span from 5\% to 23\% of the stellar surface area for 2~kG magnetic fields, and from 0.4\% to 16\% for 4~kG fields. In contrast, the super-flaring loops have IQRs ranging from 0.05\% to 1\% for 2~kG fields, and from 0.001\% to 0.1\% for 4~kG fields. Octupolar solutions are not feasible for the X-ray flares of three stars and are only partially valid for one star.

\begin{figure*}
    \centering
     \includegraphics[width=0.80\textwidth]{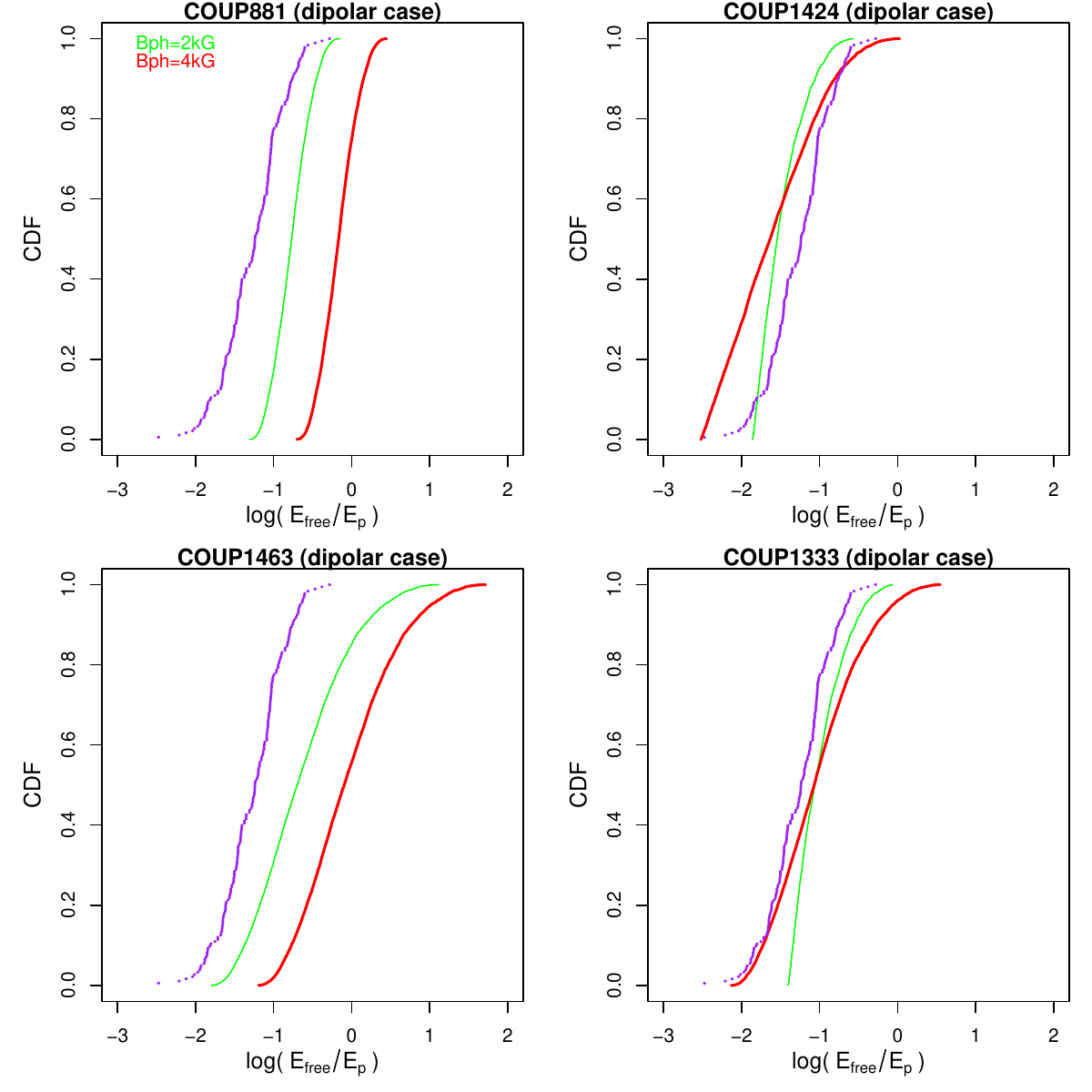}
    \caption{Cumulative distribution functions (CDFs) of the ratios of free magnetic energy to potential magnetic energy ($E_{free}/E_{p}$) in the flaring loops of the four HPF stars, shown for dipolar  magnetic loop topologies (red and green curves). These CDFs correspond to two values of photospheric magnetic field strength: 2~kG (green) and 4~kG (red). Additionally, the empirical cumulative distribution functions (ECDFs) of the ($E_{free}/E_{p}$) ratios for numerous solar M- and X-class flares, as reported by \citet{Aschwanden2014,Gupta2021}, are presented as purple curves.} \label{fig:efree_to_epot}
\end{figure*}

\begin{figure*}
    \centering
     \includegraphics[width=0.95\textwidth]{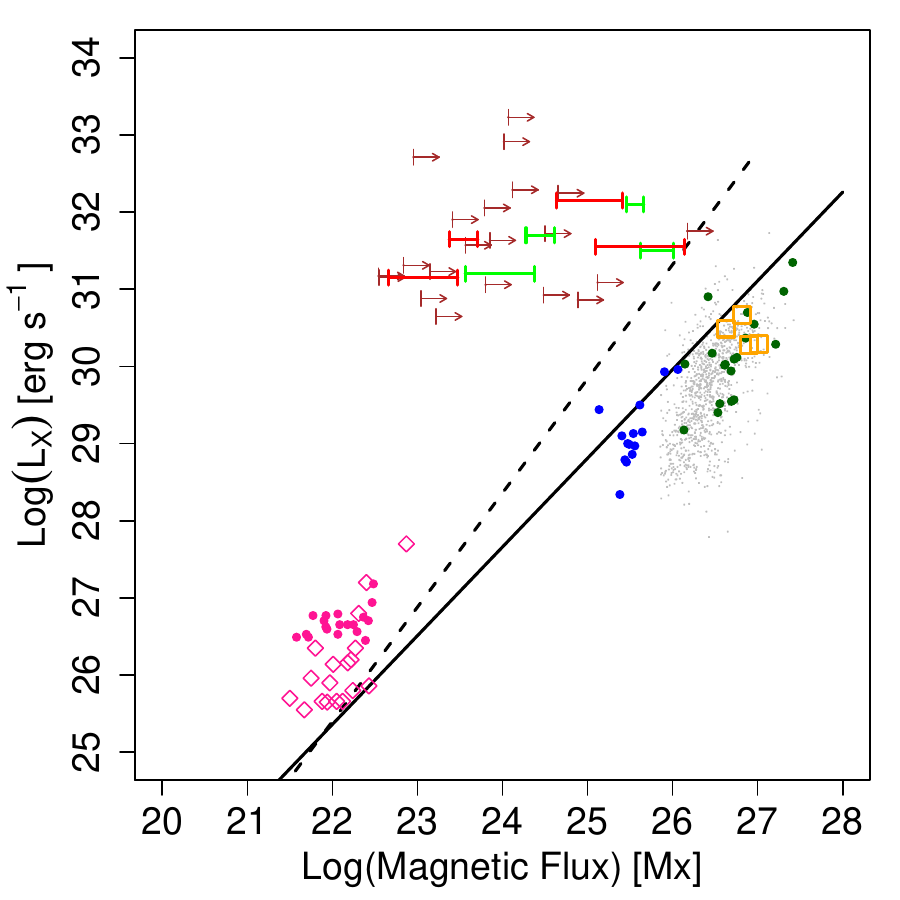}
    \caption{Zoomed-in view of X-ray luminosity ($L_X$) versus magnetic flux ($\Phi$) from Figure~\ref{fig:pevtsov_plot}, focusing on the upper-right region of the plot, where both $L_X$ and $\Phi$ are at their high ends. In addition to the elements shown in Figure~\ref{fig:pevtsov_plot}, the following new components are added: our four HPF stars, assuming their characteristic X-ray luminosities and HPF-based magnetic fluxes (Tables~\ref{tab:chandra_het_targets} and \ref{tab:mag_field_strengths}) averaged over their entire stellar surfaces (orange squares); numerous fully-convective PMS members of the IC~348, ONC, and NGC~2264 stellar clusters (gray points), with time-averaged X-ray luminosities and basic stellar properties from \citet{Getman22} and theoretical magnetic fluxes estimated using equation \ref{eqn:Bkin}; individual M-class and X-class solar flares from \citet{Su2007} (pink diamonds) and \citet{Vijayalakshmi2022} (pink points); position variants for the large X-ray flares from our four HPF stars, with dipolar solutions at 2~kG (green error bars) and 4~kG (red error bars) surface magnetic field strengths from Table~\ref{tab:mag_loop_crosssections} and with flare peak X-ray luminosity values (Table~\ref{tab:chandra_het_targets}) slightly shifted for different variants for better visibility.} \label{fig:lx_phi_zoomedin}
\end{figure*}

\subsection{Magnetic Energies} \label{sec:magnetic_energies}

As an energy consistency check, we can estimate the distributions of the ratio of free to potential magnetic energies in the flaring coronal loops of our four HPF stars by combining the previously reported empirical energy partition of large X-ray flares from young stars \citep{Flaccomio2018} with the semi-empirical simulations from \S~\ref{sec:cross_sections_flaring_loops}. These distributions can then be compared with those observed in solar M- and X-class flares \citep{Aschwanden2014,Gupta2021}.

The potential magnetic energy of a coronal magnetic loop ($E_p$) can be estimated using the equation:

\begin{equation}
E_p = 2 \int_0^{\text{LoopScale}} \frac{B(z)^2}{8 \pi} S(z) dz
\label{eqn:potential_energy_one}
\end{equation}

By combining this with equations \ref{eqn:Bloop} and \ref{eqn:flux_conservation}, the integral can be solved as:

\begin{equation}
E_p = \frac{B_{\text{phot}}^2 \, S_{\text{base}} \, R_{\star}}{4 \pi (\alpha - 1)} \left( 1 - \left( \frac{\text{LoopScale}}{R_{\star}} + 1 \right)^{1 - \alpha} \right)
\label{eqn:potential_energy_two}
\end{equation}

In addition to its potential energy, a coronal loop accumulates free magnetic energy ($E_{free}$) due to twisting and shearing motions. This free energy serves as the primary energy source for solar and stellar flare phenomena related to magnetic reconnection, including associated coronal mass ejections. The study by \citet{Flaccomio2018} on numerous powerful flares from PMS stars in the NGC~2264 star-forming region, based on simultaneous optical, infrared, and X-ray observations, suggests that the ratio of optical to X-ray flare energies for PMS flares with $E_{X,fl} > 10^{35}$~erg is approximately 5. For less powerful solar and stellar flares, this ratio is higher \citep{Woods2006, Flaccomio2018}. Assuming that the energy of a potential flare-related coronal mass ejection is comparable to the X-ray energy of a large PMS flare \citep{Moschou2019}, and considering that the bolometric radiated energy of the flare is approximately twice the optical energy \citep{Kretzschmar2011, Flaccomio2018}, we can estimate the free magnetic energy for the flares in the four HPF stars as:

\begin{equation}
E_{free} \simeq 11 \times E_{X,fl}
\label{eqn:free_energy}
\end{equation}

For valid ranges of the dipolar loop's base cross-sections (with $S_{\text base} < S_{HPF/2}$; Figure~\ref{fig:simulation_results_dipolar_case}), we estimate the ranges of the $E_{free}/E_{p}$ ratio for the large X-ray flares in the four HPF stars using equations \ref{eqn:potential_energy_two} and \ref{eqn:free_energy}. We then compare these ranges with the   $E_{free}/E_{p}$ ratios of more than 180 M- and X-class flares as tabulated in \citet{Aschwanden2014,Gupta2021}. This comparison is shown in Figure~\ref{fig:efree_to_epot}. 

The primary finding is that, for the vast majority of potential flaring coronal loop solutions in cases of super-flares from COUP 881 and COUP 1463 with $B_{phot} = 2$ kG and mega-flares from COUP 1424 and COUP 1333 with either $B_{phot} = 2$ kG or $B_{phot} = 4$ kG, the ratio of free to potential magnetic energy exhibits values consistent with those observed in M- and X-class solar flares (Figure~\ref{fig:efree_to_epot}). However, the majority of $B_{\text {phot} } = 4$~kG loop solutions for COUP~881 and COUP~1463 require significantly higher free-to-potential energy ratios than those observed in solar flares, with many showing free magnetic energy exceeding the potential magnetic energy of the loop. This finding suggests that the solutions of $B_{\text {phot} } = 4$kG and a dipolar loop configuration for COUP~881 and COUP~1463 may be energetically unreasonable.

\subsection{Positions of powerful X-ray flares on $L_X-\Phi$ diagram} \label{sec:positions_of_superflares_on_lx_phi}

To calculate the permissible ranges of magnetic fluxes at the bases of the flaring coronal loops in our four HPF stars, the 25th and 75th percentiles of the loop cross-section values, determined in \S~\ref{sec:new_loop_analyses}, are multiplied by $B_{\text{phot}}$. For the mega-flares from COUP~1333 and COUP~1424, some loop cross-section solutions (assuming $B_{\text{phot}}=2$~kG) exceed the empirical HPF-based half-sizes of the magnetic regions --- specifically, $S_{\text{base,75\%}} > S_{\text{HPF}}/2$ (see Table~\ref{tab:mag_loop_crosssections}) --- indicating that these solutions are not astrophysically viable. Consequently, these solutions for COUP~1333 and COUP~1424 are excluded from the calculation of magnetic fluxes by revising the 75th percentile of the loop cross-section and imposing a constraint: $S_{\text{base,75\%}} < S_{\text{HPF}}/2$.

Figure~\ref{fig:lx_phi_zoomedin} illustrates the resulting placement of the flare peak X-ray luminosities (Table~\ref{tab:chandra_het_targets}) as a function of these flare magnetic fluxes. These are represented with red ($B_{\text{phot}} = 4$~kG) and green ($B_{\text{phot}} = 2$~kG) error bars for the dipolar solutions.

The $B_{\text {phot}}=4$~kG dipolar-configuration solutions for the super-flares in COUP~881 and COUP~1463 were deemed potentially energetically unreasonable with respect to the solar $E_{free}/E_p$ ratio ranges (\S\ref{sec:magnetic_energies}). However, they are retained in Figure~\ref{fig:lx_phi_zoomedin} (two red error bars with $\log(\Phi) < 24$~Mx) in case the true free magnetic energies of the less-studied PMS flaring loops significantly exceed the potential magnetic energies of these loops, implying magnetic reconnection processes atypical of those observed in the current Sun.
 
Permissible PMS flare magnetic flux values span a wide 4-order of magnitude range from approximately $\Phi = 10^{22-23}$~Mx to $\Phi = 10^{26}$~Mx. This flux span is consistent with the locus of numerous PMS super- and mega-flares previously modeled by \citet{Getman2021b} (purely based on X-ray data with no direct information on surface magnetic field measurements) (brown arrows). 

The X-ray luminosities of the PMS large flare locus are 1 to 4-5 orders of magnitude higher compared to the predictions of the $L_X \propto \Phi^{1.15}$ law by \citet{Pevtsov2003} (black solid line), applicable to the time- and disk-averaged X-ray/magnetic activity of the Sun, old active stars (blue), and PMS stars (dark green). The PMS flare locus also deviates from the steeper law of \citet{Kirichenko2017}, $L_X \propto \Phi^{1.48}$ (dashed black line), mainly applicable for solar microflares and solar active regions.

Indeed, when including pure flare events, such as solar microflares and M-class flares, \citet{Kirichenko2017} report an even steeper relationship, $L_X \propto \Phi^{3}$. Analogous to this solar and PMS flare phenomenon, the X-ray luminosities of solar X-class flares from \citet{Vijayalakshmi2022}, shown in the graph as pink solid points\footnote{The X-ray luminosity values for all non-solar elements shown in Figure~\ref{fig:lx_phi_zoomedin} are provided in the $(0.5-8)$~keV band, while the X-ray luminosities for the depicted solar M- and X-class flares are in the GOES band of $(1.5-12.4)$~keV. To plot the M- and X-class flares using X-ray luminosities in the $(0.5-8)$~keV band, one can apply a correction based on WebPIMMS (\url{https://heasarc.gsfc.nasa.gov/cgi-bin/Tools/w3pimms/w3pimms.pl}), assuming optically thin thermal flare plasmas with temperatures of $T=(20-30)$~MK, typical for M- and X-class flares. With this correction, these solar flare points would shift upward by approximately (0.2-0.4)~dex on the graph.}, also deviate from the laws of \citet{Pevtsov2003} and \citet{Kirichenko2017}, though to a lesser extent than PMS flares --- by 1.5 orders of magnitude in X-ray luminosity.

A visual inspection of the relative positions of powerful PMS and solar flare loci in Figure~\ref{fig:lx_phi_zoomedin} suggests that the slope $m$ for solar and stellar flares likely falls between 1.5 and 4. However, more precise measurements will require a larger sample of powerful PMS and stellar flares.

In summary, the figure suggests that the conversion of magnetic flux to X-ray emission is more efficient in cases of more powerful stellar flares. Moreover, unusually strong stellar surface magnetic fields are not required for this efficient conversion in large PMS flares.

In Figure~\ref{fig:lx_phi_zoomedin}, we also added time- and disk-averaged X-ray luminosity and surface magnetic flux points (orange) for the 4 HPF stars, calculated from the values of $L_{X,char}$ and $\langle B \rangle \times 4 \times \pi \times R_{star}^2$ from Tables~\ref{tab:chandra_het_targets} and \ref{tab:mag_field_strengths}. Their placement with respect to other PMS stars with direct magnetic field measurements (dark green) and numerous PMS stars with theoretical magnetic flux estimates (grey points) again provides no indication of abnormal surface magnetic field strengths and magnetic fluxes during or soon after powerful X-ray flares (\S \ref{sec:typical_B_fields}).

\section{Discussion} \label{sec:discussion}

\subsection{Flare Structure Drives $L_X-\Phi$ Differences between Large PMS and Solar Flares} \label{sec:key_reason_solar_pms_differences}
On the $L_X-\Phi$ plane, large PMS flares exhibit surface magnetic fluxes and peak X-ray luminosities that are approximately 2-4~dex and 4-6~dex higher, respectively, than those of M- and X-class solar flares (Figure~\ref{fig:lx_phi_zoomedin}). This substantial offset between large PMS and solar flares can be attributed, primarily, to the significant differences in the sizes of their flaring structures.

For an explanation, as a reference, we can use the median values for a typical solar flaring loop arcade from Table~1 of \citet{Aschwanden2020}: loop scale ($\text{LoopScale} \simeq 30,000$~km), underlying surface area ($S_{\text{Solar,base}} \simeq \text{LoopScale}^2$), emitting volume ($V_{\text{Solar,flare}} \simeq \text{LoopScale}^3$), plasma temperature ($T_{\text{Solar}} \simeq 26$~MK), and density ($n_{e,\text{Solar}} \simeq 10^{10}$~cm$^{-3}$).

We compare these with the median values derived from our semi-empirical analysis of the mega-flare in COUP~1424 (\S~\ref{sec:cross_sections_flaring_loops}): a loop scale of 3 stellar radii (Figure~\ref{fig:COUP_calibration_curves}), a loop base cross-section covering 5\% of the stellar surface, i.e., $S_{\text{PMS,base}} \simeq 10^{12}$~km$^{2}$ (assuming the $B_{\text{phot}}=4$~kG dipolar solution), an emitting volume of $V_{\text{PMS,flare}} \simeq 10^{19}$~km$^{3}$ (from equation \ref{eqn:loop_scale}), plasma temperature of $T_{\text{PMS}} \simeq 70$~MK (Figure~\ref{fig:COUP_calibration_curves}), and density $n_{e,\text{PMS}} \simeq 4 \times 10^{9}$~cm$^{-3}$ (Figure~\ref{fig:COUP_calibration_curves} and equation \ref{eqn:V}).

The difference in magnetic flux between solar and PMS large flares primarily arises from the disparity in the surface areas of the flaring structures. Since the magnetic field strength in the active regions of our flaring PMS stars is comparable to that of the current Sun and non-flaring PMS stars (\S\S~\ref{sec:ZB_mesurements},\ref{sec:typical_B_fields}), the larger surface area involved in PMS flares drives the flux difference. Specifically, the magnetic flux ratio between PMS and solar flares can be expressed as $\log(\Phi_{\text{PMS,flare}}/\Phi_{\text{Solar,flare}}) = \log(S_{\text{PMS,base}} / S_{\text{Solar,base}}) \simeq 3$~dex.

The difference in peak X-ray luminosity between solar and PMS large flares can be understood through the definition of X-ray luminosity:

\begin{equation} L_{X,pk} = V n_e^2 P(T) \label{eqn
} \end{equation}

where $P(T)$ is the temperature-dependent radiative loss function for optically thin plasma \citep{Rosner1978,Sutherland1993}. The contributions from differences in temperature and density between PMS and solar flares are relatively small: $P(70~\text{MK}) / P(26~\text{MK}) \simeq 0.4$ and $(n_{e,\text{PMS}}/n_{e,\text{Solar}})^2 \simeq 0.2$. However, the large difference in the emitting volume of the flaring structures dominates the X-ray luminosity disparity. Specifically, the volume ratio $V_{\text{PMS,flare}} / V_{\text{Solar,flare}} \sim 4 \times 10^5$ drives a significant difference, leading to $\log( L_{X,pk,\text{PMS}} / L_{X,pk,\text{Solar}} ) \simeq 4.5$~dex.

This finding further supports the idea that, in many respects, large X-ray flares in PMS stars are scaled-up analogues of solar flares. 

Here are several other common features shared by both types of flares:
\begin{itemize}
    \item Currently, there is no strong empirical evidence suggesting that the properties of large PMS flares depend on the presence of protoplanetary disks, indicating that the footpoints of giant PMS flaring loops are anchored on the stellar surface \citep{Getman08b, Getman2021b, Getman2024}.
    \item Key flare-related properties --- such as X-ray emission measure ($EM$), duration ($\tau_{flare}$), coronal loop scale, and coronal plasma temperature ($T$) --- for PMS stars lie at the extreme upper ends of positively correlated relationships like $EM - T$, $\tau_{flare} - T$, and loop scale - $T$. These relationships are observed across a wide range of flares, including those from the modern Sun, active stars, and PMS stars \citep{Shibata1999, Aschwanden2008, Getman2021b}.
    \item  Both solar and PMS flares exhibit similar U-shaped density-temperature evolution patterns, suggesting analogous heating and cooling processes in flaring loops \citep{Reale1997, Reale2007,Getman08a, Getman2011}.
    \item The observed slope $\alpha$ of approximately 2, albeit with large uncertainties, in the flare energy distributions ($dN/dE_X \propto E_X^{-\alpha}$) for both solar and pre-main sequence (PMS) flares suggests that the distribution and dissipation of magnetic energy in stellar coronae might follow similar mechanisms \citep{Wolk05, Caramazza07, Colombo07, Stelzer07, Getman2021b, Kawai2022, Kowalski2024, Okamoto2021}.
    \item A significant number of semi-empirical solutions for the large PMS flares in the four HPF stars require the ratios of magnetic free energy to potential energy in the flaring loops ($E_{free}/E_{p}$) to align with the range observed in solar M- and X-class flares (\S~\ref{sec:magnetic_energies}).
    \item Both types of flares deviate from the baseline relationship on the $L_X-\Phi$ plane (\S~\ref{sec:flare_deviation_from_pevtsov}).
    \item The Neupert effect serves as an observational indicator of the two phases of energy transport --- impulsive and gradual --- in the standard model of solar flares \citep{Brown1971, Li1993, Veronig2002} and has also been observed in large PMS flares \citep{Audard2007, Getman2011, Flaccomio2018, Getman2023b}.
\end{itemize}

\subsection{Deviation of flares from the baseline $L_X-\Phi$ relationship} \label{sec:flare_deviation_from_pevtsov}

\citet{Fisher1998} discuss potential connections between the apparent baseline relationship $L_X \propto \Phi^m$ (where $m \simeq 1$) and various coronal heating models. They examine three distinct models: the nanoflare model \citep{Parker1988}, the Alfv\'{e}n wave model \citep{vanBallegooijen2011}, and the ``Minimum Current Corona'' model \citep{Longcope1996}. Their analysis suggests that the latter two models provide a better fit to the baseline relationship. Additionally, \citet{Kirichenko2017} propose that Alfv\'{e}n wave heating more accurately describes solar pre-flare active regions, whereas magnetic reconnection is more effective in explaining solar microflares.

Recent MHD simulations based on nanoflare-driven coronal heating for active regions of similar size, but with magnetic field strengths varying from 1 to 20 kG, predict a steeper X-ray-magnetic flux relationship with $m=3.4$ \citep{Zhuleku2021}. Figure~\ref{fig:lx_phi_zoomedin} shows that both powerful solar and PMS flares --- undoubtedly governed by magnetic reconnection --- exhibit significant deviations from the baseline relationship on the $L_X-\Phi^m$ plane, with a slope $m$ between 1.5 and 4.

Based on the information above, and considering that energy transfer (vertical Poynting flux, $S_z \propto B^{\beta}$) along magnetic field lines is more efficient in magnetic reconnection-driven mechanisms, such as nanoflares ($\beta = 2$), compared to Alfv\'{e}n wave heating ($\beta = 1$), it is reasonable to suggest that Alfv\'{e}n wave heating could play a significant role in sustaining the time- and disk-averaged X-ray emission levels observed on the Sun and other stars \citep{ToriumiAirapetian2022,Toriumi2022}. This process is likely characterized by an $L_X - \Phi$ relationship that is shallower than that of reconnection-driven powerful stellar flares.

It is also important to emphasize that the hypothesis, described in \S~\ref{sec:intro}, of a significantly stronger-than-a-few-kG surface magnetic field $B_{spot}$ in PMS super-mega-flaring active regions was based on the simulation results of \citet{Zhuleku2021}. However, to date, \citet{Zhuleku2021} have only considered cases with varying magnetic field strength and a constant area of the flaring active region. In light of our discovery of similar $B_{spot}$ strengths in active regions of the current Sun and PMS stars, both with and without large X-ray flares (\S~\ref{sec:typical_B_fields}), it would be valuable to conduct MHD simulations with constant $B_{spot}$ and varying active region areas. These simulations should be compared against the $L_X-\Phi$ relationship observed between powerful solar and PMS flares, as shown in Figure~\ref{fig:lx_phi_zoomedin}. \citet{Zhuleku2021} have indeed noted the possibility of performing such simulations in the future.

\subsection{The Dipole as the Preferred Magnetic Configuration for Large PMS X-ray Flares} \label{sec:magnetic_configuration}

Our finding that octupolar magnetic topology solutions are unfeasible for the X-ray flares from COUP~881, COUP~1333 and COUP~1424, and are significantly less common than dipolar solutions for the flare in COUP~1463 (\S~\ref{sec:cross_sections_flaring_loops}), indirectly supports existing empirical evidence from Zeeman-Doppler imaging data. These results indicate that the strong large-scale magnetic fields of fully convective PMS stars are primarily dipolar, but they become more complex, often involving octupolar magnetic configurations, as these stars evolve and develop radiative cores \citep{Gregory2012}. The emergence of large radiative cores triggers a transition in dynamo mechanisms from convection-driven to tachoclinal. This shift favors more intricate magnetic morphologies, which confine smaller volumes of X-ray-emitting plasma, ultimately leading to a decline in X-ray luminosity over time \citep{Gregory16,Getman22,Getman2023, Stuart2023}.   

Three-dimensional HD and MHD simulations of solar-mass PMS stars also demonstrate a decrease in the contribution of the dipole component to the stellar magnetic field throughout PMS stellar evolution \citep{Emeriau-Viard2017}.

Anelastic MHD simulations predict the formation of large polar magnetic spots that support strong large-scale dipolar fields in fully convective, rapidly rotating stars such as PMS stars and low-mass M-dwarfs \citep{Yadav2015b,Cohen2017}. The Coriolis force acting on convective rising flux tubes, deflecting them toward the poles, and/or hydrodynamical large-scale surface flows transporting magnetic fields from low to high latitudes, are likely key mechanisms contributing to the formation of these polar magnetic regions \citep{Schuessler1992,Schrijver2001,Isik2011}.

On the present-day Sun, the Coriolis force plays a crucial role in generating the tilt of bipolar sunspot groups on the solar surface, as described by Joy's law \citep[e.g.,][]{vanDriel-Gesztelyi2015, Schunker2020}. The migration of large magnetic spots towards the polar regions of PMS stars might represent a scaled-up analogue of the solar Joy's law.

By analogy with small-scale multipolar solar coronal flaring structures, the migration of strong bipolar magnetic fluxes toward and near the poles in PMS stars could be expected to trigger the non-potentiality of associated giant coronal loops, leading to significant X-ray flares. Non-potential magnetic fields may arise from energy input into potential magnetic fields due to shuffling, shearing, twisting, and merging motions driven by magnetoconvection at various scales, as well as differential rotation \citep{SteinNordlund2006, Lynch2019, Chian2023}.

\section{Conclusions} \label{sec:conclusions}

There exists an empirical universal power-law relationship between X-ray luminosity ($L_X$) and total (unsigned) surface magnetic flux ($\Phi$) applicable across solar magnetic elements, as well as time- and disk-averaged emission from the Sun, older active stars, and PMS stars. However, previous modeling of large PMS flares, based solely on X-ray data without direct measurements of magnetic fields, showed discrepancies from this universal (baseline) $L_X - \Phi$ law. Recent 3D-MHD simulations suggest that these discrepancies could be resolved if the magnetic fields in the active regions associated with powerful PMS flares reach unusually high values of 10-20~kG (\S~\ref{sec:intro}). 

To test this hypothesis, our current study employed nearly simultaneous {\it Chandra} X-ray and HET-HPF near-IR observations of young stars in the Orion Nebula, with the aim of capturing and characterizing powerful PMS X-ray flares and measuring surface magnetic field strengths during or shortly after these events. Additionally, we seek to verify and explain the positions of the observed PMS flares on the $L_X - \Phi$ plane. Four young M-type stellar members of the Orion Nebula cluster were involved in this study (\S~\ref{sec:observations}). 

This research is part of a larger multi-observatory project (MORYSEF) with the broader goals of investigating particle ejections, disk ionization, and surface magnetic fields following powerful PMS X-ray flares \citep{Getman2024}.

The magnetic intensification technique, utilizing HPF-based Ti lines in the range of $960 - 980$~nm, yielded average surface magnetic field strengths of $\langle B\rangle \simeq (2-3)$~kG in all four flaring HPF stars (\S~\ref{sec:ZB_mesurements}). These values are consistent with the \bs\ measurements of numerous PMS stars previously obtained during their likely non-flaring states (\S~\ref{sec:typical_B_fields}). These results effectively disprove our hypothesis that unusually strong surface magnetic fields occur during or after large PMS X-ray flares.

A semi-empirical flare analysis was conducted (\S\S~\ref{sec:cross_sections_flaring_loops}-\ref{sec:positions_of_superflares_on_lx_phi}) to: 1) utilize direct {\it Chandra} X-ray measurements of flare peak X-ray luminosity and flare energy, along with HPF measurements of the magnetic filling factor and average magnetic field for our four HPF stars; 2) compare the X-ray measurements against calibration curves for various previously characterized COUP PMS flares and assess typical interquartile ranges (IQR) of important flare characteristics; 3) apply Monte Carlo simulations to draw flare property values from these IQRs and analyze them through analytical equations for thermal plasma confined within dipolar or octupolar coronal magnetic loops, inferring IQR ranges of permissible surface magnetic fluxes for the flares in our four HPF stars; and 4) incorporate this flare information into the $L_X-\Phi$ plane.

Both powerful PMS flares and M/X-class solar flares exhibit deviations from the baseline $L_X-\Phi$ law, with the deviation for PMS flares being significantly more pronounced (Figure~\ref{fig:lx_phi_zoomedin}). This disparity is primarily attributed to the considerably larger volumes of PMS flaring structures compared to their solar counterparts, eliminating the need for unusually strong surface magnetic fields (\S~\ref{sec:key_reason_solar_pms_differences}). Notably, the inferred ratios of magnetic free to magnetic potential energies in PMS flaring loops are often comparable to those in solar flaring loops (\S~\ref{sec:magnetic_energies}). 

The overall deviations of solar and PMS flares from the baseline law may stem from the fact that these flares are governed by magnetic reconnection processes, while the production of baseline X-ray emission in the Sun, older active stars, and PMS stars could involve non-reconnection mechanisms, such as Alfv\'{e}n wave heating, which is less efficient in energy transfer than magnetic reconnection (\S~\ref{sec:flare_deviation_from_pevtsov}).

Our findings also indicate a preference for dipolar magnetic loop topologies over octupolar ones in our Orion PMS flares, consistent with direct Zeeman-Doppler imaging of fully convective PMS stars. Furthermore, the necessity for giant dipolar loops to generate powerful PMS X-ray flares aligns with existing MHD predictions that suggest strong dipolar loops are supported by giant polar magnetic regions in rapidly rotating, fully convective stars (\S~\ref{sec:magnetic_configuration}).

\section{Acknowledgments}
We are grateful to the anonymous referee for providing thoughtful and helpful comments that improved the manuscript. This project is supported by the SAO {\it Chandra} grant  GO3-24010X (K. Getman, Principal Investigator) and the {\it Chandra} ACIS Team contract SV4-74018 (G. Garmire \& E. Feigelson, Principal Investigators), issued by the {\it Chandra} X-ray Center, which is operated by the Smithsonian Astrophysical Observatory for and on behalf of NASA under contract NAS8-03060. The {\it Chandra} Guaranteed Time Observations (GTO) data used here and listed in \citet{Getman05} were selected by the ACIS Instrument Principal Investigator, Gordon P. Garmire, of the Huntingdon Institute for X-ray Astronomy, LLC, which is under contract to the Smithsonian Astrophysical Observatory; contract SV2-82024. O. Kochukhov acknowledges support by the Swedish Research Council (grant agreements no. 2019-03548 and 2023-03667). V.S.A. was supported by the GSFC Sellers Exoplanet Environments Collaboration (SEEC), which is funded by the NASA Planetary Science Divisions Internal Scientist Funding Model (ISFM), the NASA NNH21ZDA001N-XRP F.3 Exoplanets Research Program grant. Support for C.J.L. was provided by NASA through the NASA Hubble Fellowship grant No. HST-HF2-51535.001-A awarded by the Space Telescope Science Institute, which is operated by the Association of Universities for Research in Astronomy, Inc., for NASA, under contract NAS5-26555. S.A.D. acknowledges the M2FINDERS project from the European Research
Council (ERC) under the European Union's Horizon 2020 research and innovation programme (grant No 101018682). This paper employs a list of Chandra datasets, obtained by the Chandra X-ray Observatory, contained in~\dataset[DOI: 10.25574/cdc.285]{https://doi.org/10.25574/cdc.285}. Our results are based on observations obtained with the Habitable-zone Planet Finder Spectrograph on the Hobby-Eberly Telescope. We thank the resident astronomers and telescope operators at the HET for the execution of our observations with HPF. The Hobby-Eberly Telescope is a joint project of the University of Texas at Austin, the Pennsylvania State University, Ludwig-Maximilians-Universit\"at M\'unchen, and Georg-August Universit\"at G\"ottingen. The HET is named in honor of its principal benefactors, William P. Hobby and Robert E. Eberly.

\vspace{5mm}
\facilities{CXO, HET-HPF}

\software{R \citep{RCoreTeam20}, CIAO \citep{Fruscione2006}, {\sc SoBat} \citep{Anfinogentov2021}, {\sc Synmast} \citep{Kochukhov2007,Kochukhov2010}}


\bibliography{my_bibliography}{}
\bibliographystyle{aasjournal}

\end{document}